%% file: ttx_nlo.tex
\documentclass[a4paper,11pt]{article}
\usepackage{jheppub} 
\usepackage[T1]{fontenc}
\usepackage{booktabs}
\usepackage[dvips,table]{xcolor} 
\usepackage{xspace}
\usepackage{dcolumn}
\usepackage{hyperref}
\usepackage{caption}
\usepackage
[subrefformat=parens,position=top,skip=-15pt,margin=15pt,justification=justified,singlelinecheck=false]
{subcaption}

\newcommand{\eq}[1]{\begin{equation} #1 \end{equation}}
\newcommand{\eqa}[1]{\begin{eqnarray} #1 \end{eqnarray}}

\allowdisplaybreaks[1]

\input{macros}

\title{NLO electroweak corrections to off-shell top--antitop production with leptonic decays at the LHC}
\subheader{\today}

\author{Ansgar Denner,}
\author{Mathieu Pellen}

\affiliation{%
        Universit\"at W\"urzburg, %
        Institut f\"ur Theoretische Physik und Astrophysik, %
        Emil-Hilb-Weg 22, \linebreak %
        97074 W\"urzburg, %
        Germany%
}
\emailAdd{ansgar.denner@physik.uni-wuerzburg.de}
\emailAdd{mathieu.pellen@physik.uni-wuerzburg.de}

\abstract{For the first time the next-to-leading-order electroweak corrections to the full off-shell production of two top quarks that decay leptonically are presented.
This calculation includes all off-shell, non-resonant, and interference effects for the 6-particle phase space.
While the electroweak corrections are below one percent for the integrated cross section, they reach up to $15\%$ in the high-transverse-momentum region of distributions.
To support the results of the complete one-loop calculation, we have in addition evaluated the electroweak corrections in two different pole approximations, one requiring two on-shell top quarks and one featuring two on-shell W bosons.
While the former deviates by up to $10\%$ from the full calculation for certain distributions, the latter provides a very good description for most observables.
The increased centre-of-mass energy of the LHC makes the inclusion of electroweak corrections extremely relevant as they are particularly large in the Sudakov regime where new physics is expected to be probed.  }

\begin{document}

\maketitle
\flushbottom

\newpage

\section{Introduction}
\label{sec:introduction}

The top quark is the heaviest elementary particle known so far and
decays before it hadronises.  Its examination is of prime importance
at the Large Hadron Collider (LHC)
\cite{Chatrchyan:2012bra,Chatrchyan:2013faa,Aad:2014kva,Khachatryan:2015uqb,delDuca:2015gca,Aaboud:2016pbd}.
Therefore its production and decay rate should be computed and
measured with the highest possible precision.  In that respect, not
only next-to-leading-order (NLO) QCD
corrections~\cite{Nason:1989zy,Beenakker:1990maa,Mangano:1991jk,Frixione:1995fj,Bernreuther:2004jv,Melnikov:2009dn,Bernreuther:2010ny,Campbell:2012uf,Denner:2010jp,Bevilacqua:2010qb,Denner:2012yc,Frederix:2013gra}
including parton-shower matching
\cite{Frixione:2003ei,Frixione:2007nw,Hoeche:2014qda,Garzelli:2014dka,Campbell:2014kua,Jezo:2016ujg}
but also next-to-next-to-leading-order (NNLO) QCD
\cite{Czakon:2013goa,Czakon:2016ckf,Czakon:2016dgf}, resummation
\cite{Beneke:2009rj,Czakon:2009zw,Ahrens:2010zv,Kidonakis:2009ev,Kidonakis:2010dk},
and NLO electroweak (EW) corrections
\cite{Beenakker:1993yr,Bernreuther:2005is,Kuhn:2005it,Moretti:2006nf,Kuhn:2006vh,Bernreuther:2006vg,Hollik:2007sw,Bernreuther:2008aw,Bernreuther:2008md,Beneke:2015lwa,Pagani:2016caq}
must be considered.  The latter have so far exclusively been computed
for on-shell top quarks.  We fill this gap by computing for the first
time the EW corrections to the full off-shell production of two top
quarks that decay leptonically.  Because typical EW corrections are of
the order of the NNLO QCD ones, they must be included in any precise
analysis.  Moreover, they can grow large in particular regions of the
phase space such as for large transverse momenta.  This is
particularly relevant, as for run~II the LHC is performing at a never
accessed centre-of-mass energy.  The EW corrections are specifically
relevant in the tails of distributions where new-physics contributions
are expected to appear.  Thus, EW corrections constitute a
non-negligible Standard Model background in the phase-space regions
relevant for new-physics searches
\cite{Frederix:2007gi,Bernreuther:2015fts,Greiner:2014qna,Backovic:2015soa,Arina:2016cqj,Hespel:2016qaf}.
Improving the Standard Model predictions allows to further constrain
new-physics models or could reveal discrepancies with experimental
measurements.

In this article, the first calculation of the full NLO EW corrections
to the hadronic production of a positron, a muon, missing energy, and
two bottom-quark~jets, \ie $\fullProcess$, at the LHC is reported.
This final state is dominated by the production of a pair of top
quarks that then subsequently decay leptonically.  In particular, all
off-shell, non-resonant, and interference effects are taken into
account.  Moreover, the dominant photon-initiated process is included
for reference.

In order to support our findings we have compared the full computation
to two approximate ones.  Namely, we have also computed the EW
corrections in a double-pole approximation (DPA) with two resonant W
bosons and one with two resonant top quarks following the methods of
\citeres{Denner:2000bj,Accomando:2004de}.  This technique has been
shown to be useful in the past when computing EW corrections to
Drell--Yan processes
\cite{Wackeroth:1996hz,Baur:1998kt,Dittmaier:2001ay} as well as
di-boson production
\cite{Jadach:1996hi,Denner:1997ia,Jadach:1998tz,Beenakker:1998gr,Chapovsky:1999kv,Denner:2000bj,Bredenstein:2005zk,Billoni:2013aba}.
It has the advantage that it does not require the knowledge of the
full virtual corrections which usually constitutes the bottleneck of
this type of computations.  Nonetheless one can approximate the full
virtual corrections with an accuracy of few per cent with respect to
the leading-order (LO) contribution for many observables.  This
accuracy is often below the experimental resolution, and thus the pole
approximation is sufficient.  Recently, the EW non-factorisable
corrections needed for pole approximations have been derived in a
general form in \citere{Dittmaier:2015bfe}, and these results have
been used extensively in the present work.  We thus assess the quality
of two DPAs for the production of off-shell top quarks, which is so
far the most complicated process where it has been applied.

From a technical points of view, this computation has been made
possible thanks to two ingredients.  First the implementation of
powerful in-house multi-channel Monte Carlo program \cite{MoCaNLO}.
The second aspect is the use of the fast and reliable matrix-element
generator \recola \cite{Actis:2012qn,Actis:2016mpe} at the Born and
one-loop level.\footnote{We have used version 1.0 of \recola which is
  publicly available at http://recola.hepforge.org.}  This set-up
allows us to compute processes with a complexity equal to or higher
than the state-of-the-art NLO calculations
\cite{Denner:2015yca,Bevilacqua:2015qha,Kallweit:2015dum,Cordero:2015hem,Biedermann:2016yvs,Biedermann:2016guo}.

This article is organised as follows: in \refse{sec:calculation} the
set-up of the calculation is specified.  In particular, details about
the real (\refse{ssec:RealCorrections}) and virtual
(\refse{ssec:VirtualCorrections}) corrections are provided.  The two
DPAs considered are introduced in \refse{sec:DoublePoleApproximation},
and the checks we have performed are exposed in
\refse{sec:Validations}.  Finally, in \refse{sec:results} numerical
results are presented for a centre-of-mass energy of $\sqrt{s}=13\TeV$
at the LHC.  More specifically, in \refse{ssec:InputParameters} the
input parameters and selection cuts are specified.  The results for
integrated cross sections and distributions appear in
\refse{ssec:IntegratedCrossSection} and
\refse{ssec:DifferentialDistributions}, respectively.  In
\refse{sec:ComparisonDPA} the full calculation and the DPAs are
compared both at the level of the total cross section and of
distributions.  Our concluding remarks appear in
\refse{sec:Conclusions}.

\section{Details of the calculation}
\label{sec:calculation}

\begin{figure}
        \newcommand{\myframebox}{\framebox}
        \renewcommand{\myframebox}{\relax}
        \setlength{\parskip}{-10pt}
        \captionsetup[subfigure]{margin=5pt}
        \begin{subfigure}{0.32\linewidth}
                \myframebox{
                        \includegraphics[width=\linewidth]{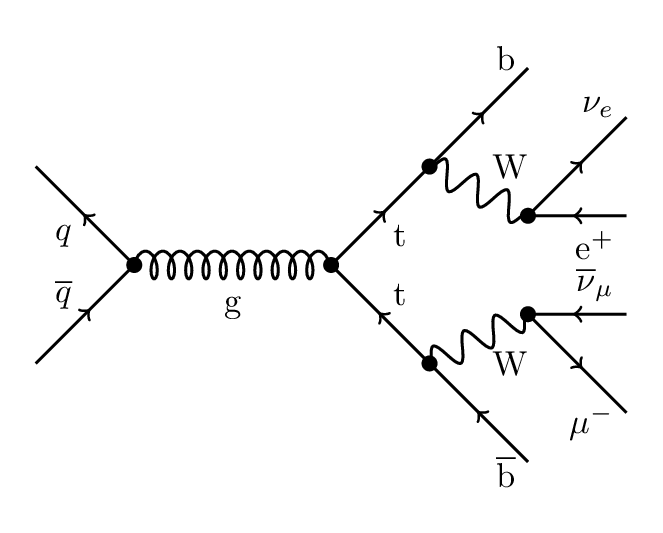}
                }
                \label{fig:born_2tops_gg_tchannel} 
        \end{subfigure}
        \begin{subfigure}{0.37\linewidth}
                \myframebox{
                        \includegraphics[width=\linewidth]{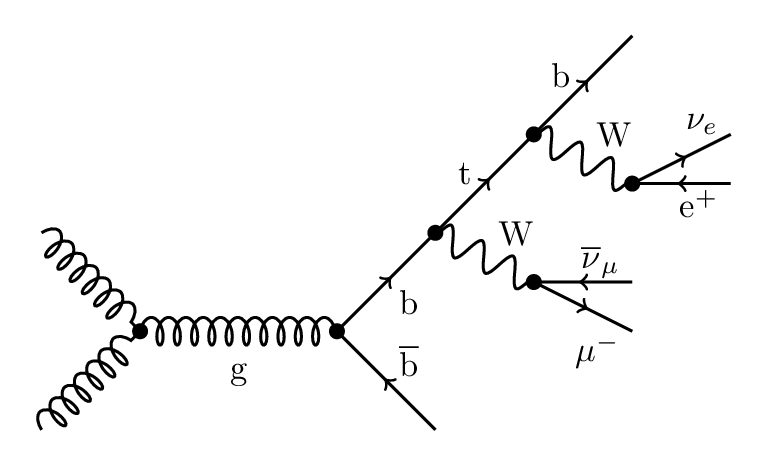}
                }
                \label{fig:born_2tops_gg_schannel} 
        \end{subfigure}
        \begin{subfigure}{0.27\linewidth}
                \myframebox{
                        \includegraphics[width=\linewidth]{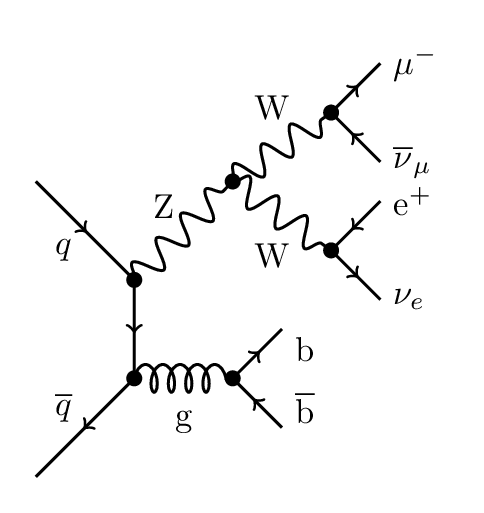}
                }
                \label{fig:born_2tops_uxu_schannel} 
        \end{subfigure}

        \caption{\label{fig:lo_tree_feynman_diagrams}%
          Representative tree-level Feynman diagrams with two (left),
          one (middle) and no (right) top-quark resonances.  }
\end{figure}

In this article, the EW corrections to the full hadronic process
\begin{equation}\label{eqn:full_process}
        \fullProcess 
\end{equation}
are considered.  The tree-level matrix element squared contributes at
the order $\order{\alphas^2\alpha^{4}}$.  The EW corrections to this
process comprise all possible corrections of the order
$\order{\alphas^2\alpha^{5}}$.  Moreover, the tree-level $\gamma {\rm
  g}$ contributions which are of the order $\order{\alphas\alpha^{5}}$
have been included for reference.  In principle one should also take
into account the QCD corrections to these contributions which are of
the order $\order{\alphas^2\alpha^{5}}$.  Since the $\gamma {\rm g}$
channel contributes only at the level of a per cent, the corresponding
QCD corrections, which form a gauge-independent subset, are expected
to be at the per-mille level with respect to the LO of the process
\eqref{eqn:full_process} and have therefore been neglected.  In the
present calculation all interferences, resonant, non-resonant, and
off-shell effects of the top quarks as well as the gauge bosons are
taken into account.  In \reffi{fig:lo_tree_feynman_diagrams} some
diagrams for two, one, and no resonant top quark(s) are displayed.
Note that the quark-mixing matrix has been assumed to be diagonal.
Moreover, the contributions originating from the bottom-quark parton
distribution function (PDF) have been neglected.

The calculation is performed with the in-house multi-channel Monte
Carlo program {\sc MoCaNLO} \cite{MoCaNLO} which has proven to be
particularly suited for complicated processes with high multiplicity
\cite{Denner:2015yca}.  It uses phase-space mappings similar to those
of Refs.~\cite{Berends:1994pv,Denner:1999gp,Dittmaier:2002ap}.
Infrared (IR) singularities in the real contributions are handled by
the dipole subtraction method
\cite{Catani:1996vz,Dittmaier:1999mb,Catani:2002hc,Phaf:2001gc}
implemented in a general manner for both QCD and QED.  The
matrix-element generator \recola-1.0 \cite{Actis:2012qn,Actis:2016mpe}
and the loop-integral library \collier-1.0\footnote{We have used the
  public version of \collier that can be found at
  http://collier.hepforge.org.}~\cite{Denner:2014gla,Denner:2016kdg}
have been linked to the Monte Carlo code.  They are used for the
computation of all tree and one-loop amplitudes and all ingredients
needed for the subtraction terms such as colour- and spin-correlated
squared amplitudes.  The calculation presented here is similar to
those for $\Pp\Pp\to\Pe^+\nu_\Pe \mu^-\bar{\nu}_\mu\Pb\bar{\Pb} \PH$
in~\citere{Denner:2015yca} and $\Pp\Pp\to\Pe^+\nu_\Pe
\mu^-\bar{\nu}_\mu\Pb\bar{\Pb}$ in \citere{Denner:2012yc} in many
respects.  In particular, the selection cuts considered are almost
identical, and the same computer programs have been used as in
\citere{Denner:2015yca}.

\subsection{Real corrections}
\label{ssec:RealCorrections}

\begin{figure}
        \newcommand{\myframebox}{\framebox}
        \renewcommand{\myframebox}{\relax}
        \setlength{\parskip}{-10pt}
        \captionsetup[subfigure]{margin=5pt}
        \begin{subfigure}{0.48\linewidth}
                \myframebox{
                        \includegraphics[width=\linewidth]{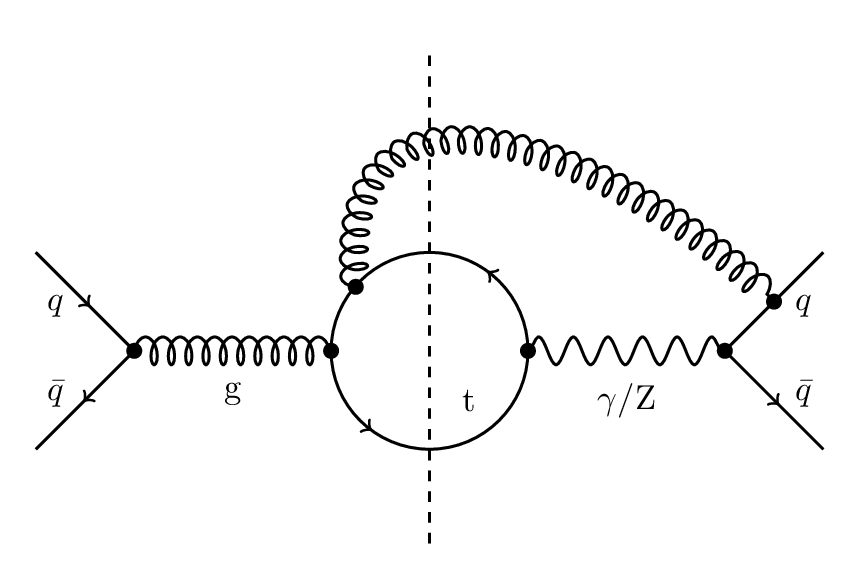}
                }
        \end{subfigure}
        \begin{subfigure}{0.48\linewidth}
                \myframebox{
                        \includegraphics[width=\linewidth]{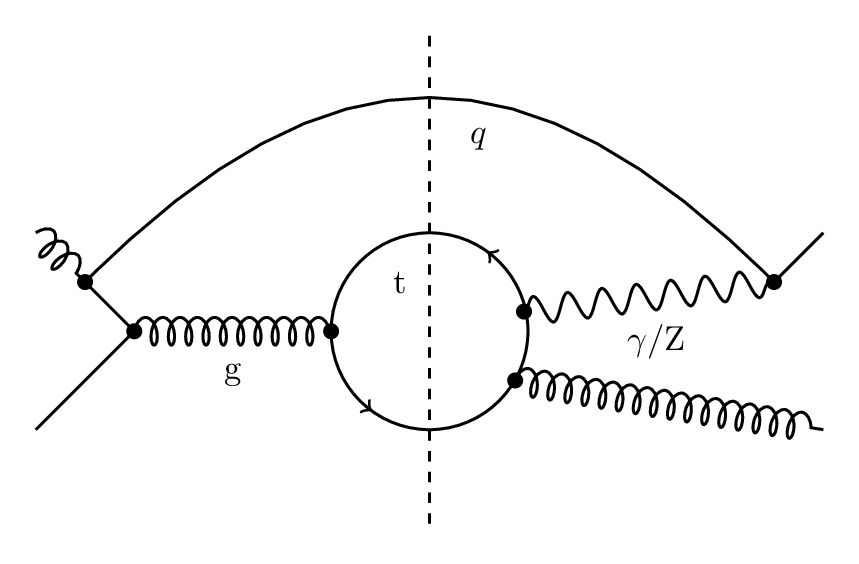}
                }
        \end{subfigure}

        \caption{\label{fig:real_int_feynman_diagrams}%
          Representative real Feynman diagrams squared which feature
          interference between QCD and EW tree-level diagrams.  Only
          the top quarks are represented as the inclusion of their
          decay products does not alter the discussion.  }
\end{figure}

The real corrections comprise all the real-radiation contributions of
order $\order{\alphas^2\alpha^{5}}$ to the process
\eqref{eqn:full_process}.  The first type of real corrections is due
to photons radiated from any of the charged particles involved in the
tree-level process $\fullProcess$.  As we are aiming at the complete
$\order{\alphas^2\alpha^{5}}$ corrections, interferences of a QCD
production of the pair of top quarks and a gluon with its EW
counterpart in the ${ q \bar{q}}$ channel must be taken into account.
Note that because of the colour structure, the only non-zero
contributions are the interferences between initial- and final-state
radiation diagrams.  This is exemplified on the left-hand side
of~\reffi{fig:real_int_feynman_diagrams}.  The squared Feynman
diagrams are represented in the figure with on-shell top quarks in
order to simplify the representation, but the final state considered
in the calculation does not involve two on-shell top quarks but rather
four leptons and two bottom-quark jets.  In the same manner, another
type of interference appears, namely the interference in the ${ q g}$
or ${ \bar{q} g}$ channel as shown on the right-hand side
of~\reffi{fig:real_int_feynman_diagrams}.

For the treatment of the IR singularities, the Catani--Seymour
subtraction formalism \cite{Catani:1996vz,Catani:2002hc} has been used
for QCD and its extension to QED \cite{Dittmaier:1999mb}.  The QCD
singularities from collinear initial-state splittings have been
absorbed in the PDFs using the $\overline{\text{MS}}$ factorisation
scheme.  The NNPDF collaboration \cite{Ball:2013hta} states that the
$\mathrm{NNPDF23}\_\mathrm{nlo}\_\mathrm{as}\_0119\_\mathrm{qed}$ PDF
sets can be used in any reasonable factorisation scheme for QED, as
the QED evolution is taken into account at leading-logarithmic level.
Nonetheless the use of different factorisation schemes differs by
next-to-leading logarithms, and the perturbative expansion can show
better convergence in certain schemes
\cite{Diener:2005me,Dittmaier:2009cr}.  For this reason, the EW
collinear initial-state splittings have been handled using the DIS
factorisation scheme.  The difference between the two schemes turned
out to be below the integration error at the total cross-section
level.  Even if noticeable (around $1 \%$) for the quark-induced
channels, the difference is negligible for the total cross section as
the gg channel (which does not feature initial-state photon radiation)
is dominant.  Note finally that all the squared amplitudes for the
real-correction sub-processes as well as the colour- and
spin-correlated squared amplitudes have been obtained from the
computer code \recola \cite{Actis:2012qn,Actis:2016mpe}.

\subsection{Virtual corrections}
\label{ssec:VirtualCorrections}

\begin{figure}
        \newcommand{\myframebox}{\framebox}
        \renewcommand{\myframebox}{\relax}
        \setlength{\parskip}{-10pt}
        \captionsetup[subfigure]{margin=5pt}
        \begin{subfigure}{0.48\linewidth}
                \myframebox{
                        \includegraphics[width=\linewidth]{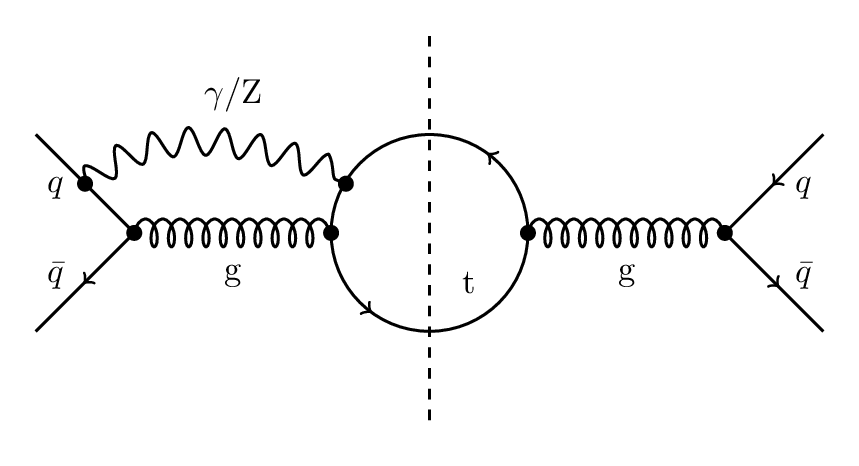}
                }
        \end{subfigure}
        \begin{subfigure}{0.48\linewidth}
                \myframebox{
                        \includegraphics[width=\linewidth]{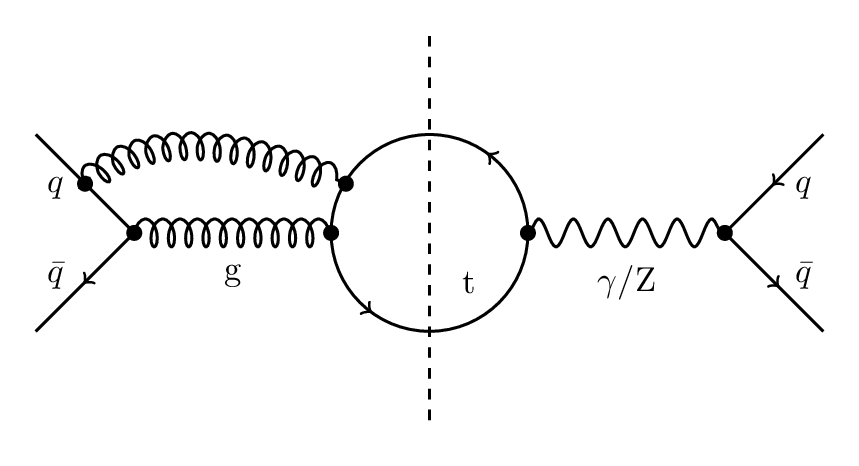}
                }
        \end{subfigure}
        \caption{\label{fig:loop_int_feynman_diagrams}%
          Representative one-loop Feynman diagrams squared.  The
          diagram on the left-hand side represents an EW correction to
          the QCD process. It can also be interpreted as a QCD
          correction to the EW amplitude interfered with the QCD
          amplitude.  The right-hand side shows a QCD correction to
          the QCD amplitude interfered with the EW amplitude.  Only
          the top quarks are represented as the inclusion of their
          decay products does not alter the discussion.  }
\end{figure}

\begin{figure}
        \newcommand{\myframebox}{\framebox}
        \renewcommand{\myframebox}{\relax}
        \setlength{\parskip}{-10pt}
        \captionsetup[subfigure]{margin=0pt}
        \begin{subfigure}{0.43\linewidth}
                \myframebox{
                        \includegraphics[width=\linewidth]{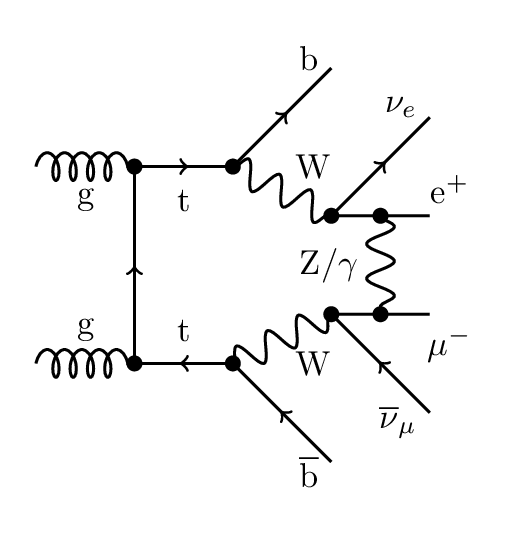}
                }
                \label{fig:virt_2tops_gg_tchannel} 
        \end{subfigure}
        \begin{subfigure}{0.51\linewidth}
                \myframebox{
                        \includegraphics[width=\linewidth]{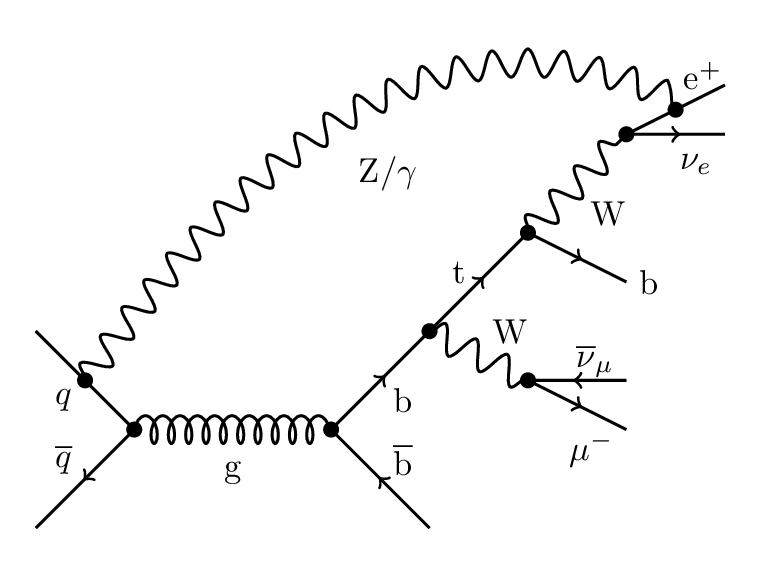}
                }
                \label{fig:virt_2tops_gg_schannel} 
        \end{subfigure}
        \caption{\label{fig:nlo_loop_feynman_diagrams}%
                Representative octagon and heptagon one-loop Feynman
                diagrams.
        }
\end{figure}

As for the real corrections, there are two types of virtual
corrections.  The first type results from the insertion of an EW
particle anywhere in the tree-level amplitude.  In the $\bar{q}q$
channel, a second type originates from the insertion of a gluon in the
QCD-mediated tree-level amplitude which is then interfered with the EW
tree-level amplitude.  These two types of corrections are depicted
in~\reffi{fig:loop_int_feynman_diagrams}.  Again only the two top
quarks and not their decay products are represented to simplify the
discussion.  Some exemplary diagrams of the most complicated loop
amplitudes (7- and 8-point functions) are depicted
in~\reffi{fig:nlo_loop_feynman_diagrams}.  The virtual corrections
have been computed in the 't~Hooft--Feynman gauge in dimensional
regularisation using the matrix-element generator \recola
\cite{Actis:2012qn,Actis:2016mpe} as well as the library \collier
\cite{Denner:2014gla,Denner:2016kdg}, which is used to calculate the
one-loop scalar
\cite{'tHooft:1978xw,Beenakker:1988jr,Dittmaier:2003bc,Denner:2010tr}
and tensor integrals
\cite{Passarino:1978jh,Denner:2002ii,Denner:2005nn} numerically.


All resonant massive particles, \ie top quarks, Z bosons and W bosons,
are treated in the complex-mass scheme
\cite{Denner:1999gp,Denner:2005fg,Denner:2006ic}.  Accordingly, the
masses of the unstable particles as well as the weak mixing angle are
consistently treated as complex quantities,
\begin{equation}
\label{e:WZcms}
 \overline{M}^2_\PW=\MW^2-\ri\MW\Gamma_\PW,\qquad
 \overline{M}^2_\PZ=\MZ^2-\ri\MZ\Gamma_\PZ, \qquad \text{and} \qquad
\cos\theta_{\mathrm{w}}=\frac{ \overline{M}_\PW}{ \overline{M}_\PZ}.
\end{equation}

\subsection{Double-pole approximation}
\label{sec:DoublePoleApproximation}

\subsubsection*{Generalities}

The dominant contributions to the process $\Pp\Pp\to\Pe^+\nu_\Pe
\mu^-\bar{\nu}_\mu\Pb\bar{\Pb}$ result from the production of two top
quarks that subsequently decay into bottom quarks and W bosons, which
in turn decay into lepton--neutrino pairs.  The simplest approximation
is thus to require two on-shell top quarks and two on-shell W bosons.
However, demanding just two on-shell top quarks is not much more
complicated, since each decaying top quark gives rise to a W boson
anyhow.  Requiring in turn only two on-shell W bosons, will thus
include also all contributions with resonant top quarks, but in
addition also all contributions with one resonant top quark.

Calculating the NLO corrections to a process with intermediate
on-shell particles implies to include the corrections to their
production and decay.  The on-shell approximation does not include
off-shell effects as well as virtual corrections that link the
production part and the decay parts or different decay parts.  Such
corrections should be of the order $\order{\Gamma_i/M_i}$
\cite{Fadin:1993kt,Fadin:1993dz,Melnikov:1993np} if the decay products
are treated inclusively and the resonant contributions dominate.  Here
$\Gamma_i$ and $M_i$ are the width and the mass of the resonant
particles, respectively.  Off-shell effects of the resonant particles
can be taken into account by using the pole approximation. In this
case, the resonant propagators are fully included, while the rest of
the matrix element is expanded about the resonance poles. Moreover,
spin correlations between production and decay can be included easily.

\begin{figure}
        \newcommand{\myframebox}{\framebox}
        \renewcommand{\myframebox}{\relax}
        \setlength{\parskip}{-10pt}
        \captionsetup[subfigure]{margin=5pt}
        \begin{subfigure}{0.48\linewidth}
                \myframebox{
                        \includegraphics[width=\linewidth]{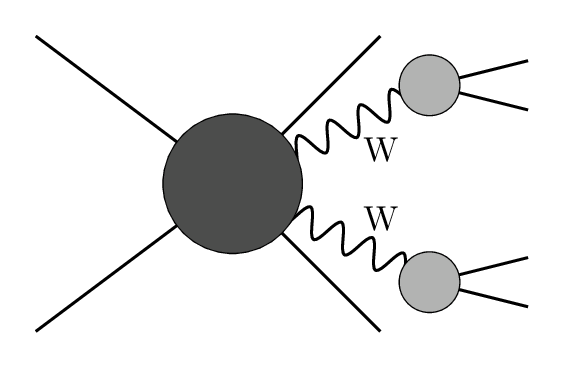}
                }
        \end{subfigure} 
        \begin{subfigure}{0.48\linewidth}
                \myframebox{
                        \includegraphics[width=\linewidth]{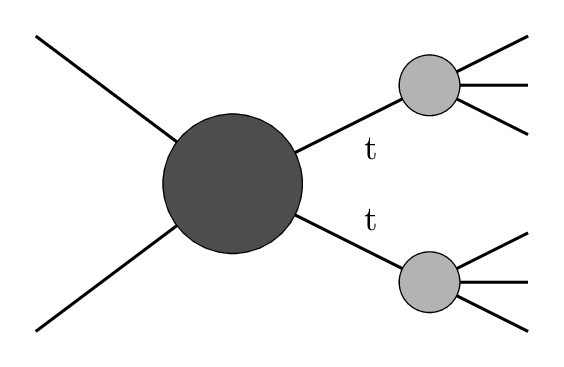}
                }
        \end{subfigure}
        \caption{\label{DPA_diagrams}%
          Schematic representation of the two DPAs.  On the left-hand
          side the two W bosons are projected on shell, while on the
          right-hand side the two top quarks are projected on shell.}
\end{figure}

We have studied%
\footnote{We have not considered a pole approximation requiring
  simultaneously two resonant top quarks and two resonant W bosons.
  First of all, requiring only a pair of resonant particles
  constitutes a better approximation.  On the other hand,
  \citere{Dittmaier:2015bfe} does not provide results for resonances
  that are part of cascade decays.}  two different DPAs for the
process \eqref{eqn:full_process} graphically represented in
\reffi{DPA_diagrams}: In one case, we require two resonant W~bosons
and in the second case two resonant top quarks.  In order to ensure
gauge invariance, the momenta of the resonant particles entering the
matrix elements have to be projected on shell. On the other hand, in
the phase space and in the propagators of the resonant particles
off-shell momenta are used.  In the DPA, as in any pole approximation,
two different kinds of corrections appear, factorisable and
non-factorisable corrections.

The factorisable virtual corrections can be uniquely attributed either
to the production of the resonant particles or to their decays.  Thus,
the diagrams displayed in~\reffi{fig:nlo_loop_feynman_diagrams} are,
for example, not included in the set of factorisable virtual
corrections.  Using the notation of \citere{Dittmaier:2015bfe} for a
pole approximation of $r$ resonances ($r=2$ for a DPA), the latter can
be written as
\begin{align}
        \mathcal{M}_{\mathrm{virt,fact,PA}} & =
        \begin{aligned}[t]
                & \sum_{\lambda_1,\ldots,\lambda_r} \left( \prod^r_{i=1} \frac{1}{K_i}\right) \Bigg[ \mathcal{M}^{I \to N,\overline{R}}_{\mathrm{virt}} \prod^r_{j=1} \mathcal{M}^{j \to R_j}_{\mathrm{LO}} \nonumber
        \end{aligned} \\
        &\quad{} +
        \begin{aligned}[t]
                &  \mathcal{M}^{I \to N,\overline{R}}_{\mathrm{LO}} \sum^r_{k=1} \mathcal{M}^{k \to R_k}_{\mathrm{virt}}
                           \prod^r_{j \neq k} \mathcal{M}^{j \to R_j}_{\mathrm{LO}} \Bigg]_{ \left\{ \overline{k}^2_l \to {\widehat{\overline{k}}}^2_l = M^2_l \right\}_{l \in \overline{R}} } , \\
        \end{aligned}
\end{align}
where $K_i = \overline{k}^2_i - \overline{M}^2_i$ is the propagator of
the resonant particle $i$, with complex mass squared $\overline{M}^2_i
= M^2_i - \text{i} M_i \Gamma_i$.
%
The on-shell projection denoted by $\left\{ \overline{k}^2_l \to
  {\widehat{\overline{k}}}^2_l = M^2_l \right\}$ is applied everywhere
in the matrix element but in the resonant propagators $K_i$.  The
indices $I$, $\overline{R}$, $R_i$ and $N$ denote the ensembles of
initial particles, resonant particles, decay products of the resonant
particle $i$, and the final-state particles not resulting from the
decay of a resonant particle.  The polarisations of the resonances are
represented by $\lambda_i$.  Alternatively, the factorisable
corrections can be obtained by selecting all Feynman diagrams for the
complete process that contain the specified $r$ resonances of the set
$\overline{R}$. Using this approach, the factorisable corrections can
be generated with the computer code \recola, which allows to select
contributions featuring resonances at both LO and NLO.

The factorisable corrections constitute a gauge-invariant subset
\cite{Stuart:1991xk,Veltman:1992tm,Aeppli:1993cb}.  As virtual
corrections, they are not IR finite in the presence of external
charged particles.  Moreover, taking the on-shell limit of the momenta
of the resonant particles introduces additional artificial IR
singularities from charged resonances.  For example, a photon exchange
between a W boson and the attached bottom quark leads to such an
artificial IR singularity, if the W~boson is projected on shell.

The virtual non-factorisable corrections arise only from diagrams
where a photon (or a gluon) is exchanged in the loop
\cite{Beenakker:1997ir,Denner:1997ia}.  On the one hand, they result
from manifestly non-factorisable diagrams, \ie diagrams that do not
split into production and decay parts by cutting only the resonant
lines, as for example those depicted in
\reffi{fig:nlo_loop_feynman_diagrams}.  On the other hand, they also
include contributions from factorisable diagrams.  The latter are
caused by IR singularities of on-shell resonances. They are obtained
by taking the factorisable diagrams, where the IR singularities
related to the resonant particles are regularised by the finite decay
widths and subtracting these contributions for zero decay width, which
contains the artificial IR-divergent piece mentioned previously.  In
general, the non-factorisable corrections factorise from the LO matrix
element and can be written in the form

\eq{2 \mathrm{Re} \left\{ \mathcal{M}^*_{\mathrm{LO,PA}} \mathcal{M}_{\mathrm{virt,nfact,PA}} \right\} = |\mathcal{M}_{\mathrm{LO,PA}}|^2 \delta_{\mathrm{nfact}} .}

In order to cancel the IR singularities in the virtual corrections,
one has to apply the on-shell projection to the terms containing the
$I$~operator in the integrated dipole contribution in the same way as
for the factorisable and non-factorisable contributions.  The $P$- and
$K$-operator terms, on the other hand, are evaluated with the
off-shell kinematics like the real corrections.  This introduces a
mismatch, which is of the order of the intrinsic error of the DPA.
Note that for the LO and all real contributions no pole approximation
is applied~\cite{Denner:2000bj,Accomando:2004de}.

As mentioned above, in the case of top-quark pair production the
 ${ q \bar{q}}$  channel has two kinds of virtual NLO contributions:
the EW loop corrections to the QCD-mediated process and the
interference of the QCD-mediated one-loop amplitude with the EW tree
amplitude.  Both contributions are connected by IR divergences and we
call the latter {\em{interference contributions}} in the following.
Thus, besides applying the DPA to the EW loop corrections of
the QCD-mediated process, we must also adopt the DPA for the second
type of corrections.  Then, also the corresponding $I$ operator has to be
evaluated with on-shell-projected kinematics.

Following the notations of \citere{Dittmaier:2015bfe}, all invariants used in the equations below are defined as:
\eqa{s &=& \left( \sum_{i\in I} p_i \right)^2, \nonumber \\  
s_{ij} &=& \left( k_i + k_j \right)^2, \qquad i,j \in I \cup F, \nonumber \\
\overline{s}_{ij} &=& \left( \overline{k}_i + k_j \right)^2, \qquad i \in \overline{R}, \qquad j \in I \cup F, \nonumber \\
\widetilde{s}_{ij} &=& \left( \overline{k}_i - k_j \right)^2, \qquad i \in \overline{R}, \qquad j \in I \cup F, \nonumber \\
\overline{\overline{s}}_{ij} &=& \left( \overline{k}_i - \overline{k}_j \right)^2, \qquad i,j \in \overline{R} ,}
where the momenta $p_i$, $k_i$ and $\overline{k}_i$ are the momenta of
the incoming, outgoing and resonant particles, respectively.  Here,
$F$ constitutes the ensemble of all the final-state particles.

\subsubsection*{Double-pole approximation for $\text{W}^+$ and
  $\text{W}^-$ bosons}
\label{sec:WWDPA}

We first discuss the DPA for two W~bosons.  In order not to shift the
top resonances, we have chosen an on-shell projection that leaves the
momenta and thus the invariants of the top quarks untouched.  Since
the 
$\PW^+$~boson 
is projected on its mass shell, one necessarily obtains:
\begin{equation}
        p_{\rm t} = p_{\rm b} + {\overline{p}}_{{\rm W}^+} = p_{\rm b} + p_{{\rm e}^+} + p_{\overline{\nu}_e} = \widehat{p}_{\rm b} + {\widehat{\overline{p}}}_{{\rm W}^+} ,
\end{equation}
where $\overline{p}$ and $\widehat{p}$ denote the four-momenta of the
resonant and the projected particles, respectively.
This leads to \cite{Dittmaier:2015bfe}
\begin{equation}
 \widehat{p}_{\rm b} = p_{\rm b} \frac{ p^2_{\rm t} - m^2_{\rm W} }{ 2 p_{\rm b} \cdot p_{\rm t}} \qquad \mathrm{and} \qquad {\widehat{\overline{p}}}_{{\rm W}^+} = p_{\rm t} - \widehat{p}_{\rm b}.
\end{equation}
In the same manner, the decay products of the resonant $\PW^+$ boson can be
written as
\begin{equation}
 \widehat{p}_{{\rm e}^+} = p_{{\rm e}^+} \frac{m^2_{\rm W}}{ 2 p_{{\rm e}^+} \cdot {\widehat{\overline{p}}}_{{\rm W}^+} } \qquad \mathrm{and} \qquad \widehat{p}_{{\nu}_e} = {\widehat{\overline{p}}}_{{\rm W}^+} - \widehat{p}_{{\rm e}^+} .
\end{equation}
The kinematic projection for the 
$\PW^-$~resonance 
is obtained by renaming the involved particles.

For the process ${\rm u \overline{u}} \to\Pe^+\nu_\Pe \Pb
\mu^-\bar{\nu}_\mu \bar{\Pb}$, the decay products of the ${\rm W}^+$
and ${\rm W}^-$~bosons are $\Pe^+\nu_\Pe$ and $\mu^-\bar{\nu}_\mu$,
respectively.  The final-state particles not resulting from a decay
are the two bottom quarks.  In the compact notation of
\citere{Dittmaier:2015bfe} this reads:
\begin{equation}
\label{eq:ensemblesNFC}
I = \{1,2\}, \qquad R_1=\{3,4\}, \qquad R_2=\{6,7\}, \qquad
  \text{and} \qquad N=\{5,8\} .
\end{equation}
The conventions for the sign factors and charges are
\begin{equation}
\label{eq:sigmaNFC}
\sigma_1 = - \sigma_2 = 1, \qquad \sigma_3 = \sigma_7 = \sigma_8 =
  1, \qquad \sigma_4 = \sigma_5 = \sigma_6 = -1 ,
\end{equation}
and
\eq{Q_1=Q_2= \frac23, \qquad Q_3=Q_6=-1 , \qquad Q_4=Q_7=0  , \qquad
  Q_5=Q_8=-\frac13 .
\label{eq:chargesNFC}
}
The results for the gluon--gluon channel are obtained upon setting
$Q_{1/2} = 0$.

Owing to the fact that the ensemble $N \cup I$ contains only pairs of
particles with opposite charges, the expression for
$\delta_{\mathrm{nfact}}$ simplifies to:
\begin{align}
        \delta_{\mathrm{nfact}} & =
        \begin{aligned}[t]
                &  - \sum_{a \in R_1} \sum_{b \in R_2} \sigma_a \sigma_b Q_a Q_b \frac{\alpha}{\pi} \mathrm{Re} \left\{ \Delta\left(i=1, a; j=2, b\right) \right\} \nonumber
        \end{aligned} \\
        &\quad{} -
        \begin{aligned}[t]
                & \sum_{i = 1}^2 \sum_{a \in R_i} \sum_{b \in N \cup I} \sigma_a \sigma_b Q_a Q_b \frac{\alpha}{\pi} \mathrm{Re} \left\{ \Delta_{\mathrm{xf}} \left(i, a; b\right) + \Delta_{\mathrm{xm}}\left(i; b\right) \right\} . \\
        \end{aligned}
\label{eq:deltanf}
\end{align}
The different contributions read:
\eq{ \Delta \left(i,a;j,b \right) = \Delta_{\mathrm{mm}} \left( i,j\right) + \Delta_{\mathrm{mf}} \left(i,a;j,b \right) + \Delta_{\mathrm{mm'}}\left(i,j\right) + \Delta_{\mathrm{mf'}} \left(i,a;j,b \right) + \Delta_{\mathrm{ff'}}\left(i, a; j, b\right) }
and are further decomposed as
\eqa{\Delta_{\mathrm{mm}} \left( i,j\right) &=& \Delta'_{\mathrm{mm}}
  \left( i\right) + \Delta'_{\mathrm{mm}} \left(j \right), \nonumber \\  
\Delta_{\mathrm{mf}} \left(i,a;j,b \right) &=& \Delta'_{\mathrm{mf}} \left(i,a \right) + \Delta'_{\mathrm{mf}} \left(j,b \right), \nonumber\\
\Delta_{\mathrm{mf'}} \left(i,a;j,b \right) &=& \Delta'_{\mathrm{mf'}}
\left(i;j,b\right) + \Delta'_{\mathrm{mf'}} \left(j;i,a \right) .
\label{eq:delta_i_nf}}
The explicit expressions for the various contributions in terms of
scalar integrals can be found in \citere{Dittmaier:2015bfe} and have
been reproduced for completeness in \refapp{sec:Appendix}.

As stated above, the pole approximation should also be applied to the
interference contributions.  Since we only consider leptonic
decays of the W~bosons, there are no QCD corrections that link
production and decay, and thus no non-factorisable interference
contributions appear for the DPA applied to the W bosons.  Nonetheless 
factorisable corrections 
of interference type exist.

Finally, note that as the width of the W boson is assumed to be zero
everywhere except in their resonant propagators, we also set the width
of the Z boson to zero.  This avoids artificially large higher-order
terms in the calculation of the complex weak mixing angle.

\subsubsection*{Double-pole approximation for t and \boldmath
  $\overline{\rm t}$ quarks}

Next we discuss the DPA for two top quarks.  We use the on-shell
projection introduced in \citere{Denner:2014wka} and reproduce it here
for completeness.  In general, one can enforce a projection of two
momenta $p_1$ and $p_2$ such that they fulfil $p_1 +p_2 =
\widehat{p}_1 + \widehat{p}_2$ with $\widehat{p}_1^2 = m_1^2$ and
$\widehat{p}_2^2 = m_2^2$, where the masses $m_1$ and $ m_2$ are
not necessarily the physical masses.  The projected momenta read:
\eq{ \label{tt1}
\widehat{p}_1 = \xi p_1 + \eta p_2, \qquad \widehat{p}_2 = (1-\xi) p_1 + (1-\eta) p_2 .}
The constants $\xi$ and $\eta$ are obtained by solving the quadratic
equation
\eqa{
\label{tt2}
0 &=& \eta^2 \left[p_1^2 p_2 - p_2^2 p_1 + \left( p_1 p_2 \right) \left( p_2 - p_1\right) \right]^2 \nonumber \\
&& {}+ \eta \left[ \left(p_1 +p_2 \right)^2 + \widehat{p}^2_1 - \widehat{p}^2_2 \right] \left[ \left(p_1 p_2 \right)^2 - p_1^2 p^2_2 \right]  \nonumber \\
&& \times \frac14 \left[ \left(p_1+p_2\right)^2 + \widehat{p}^2_1 - \widehat{p}^2_2 \right]^2 p_1^2 - \left(p_1^2 + p_1 p_2 \right)^2 \widehat{p}^2_1 }
and using
\eq{\label{tt3}
\xi = \frac{\left( \left(p_1 + p_2 \right)^2 + \widehat{p}^2_1 - \widehat{p}^2_2 \right) - 2 \eta \left( p_2^2 + p_1 p_2 \right)}{2 \left(p_1^2 + p_1 p_2\right)} .}

For the projection of the two top quarks, the only replacements needed are $p_1 \to \overline{p}_{\rm t}$, $p_2 \to \overline{p}_{\overline{{\rm t}}}$ and $\widehat{p}^2_1 = \widehat{p}^2_2 = m_{\rm t}^2$.
\sloppypar
Defining
\eq{p'_{\rm b} = \widehat{\overline{p}}_{\rm t} - p_{{\rm W}^+} ,}
it is possible to obtain $\widehat{p}_{\rm b}$ and $\widehat{p}_{{\rm
    W}^+}$ using  Eqs.~\eqref{tt1}--\eqref{tt3} 
upon performing the 
replacements 
$p_1 \to p'_{\rm b}$ and $p_2 \to
p_{{\rm W}^+}$.  The projected invariants are defined as
$\widehat{p}^2_1 = 0$ and $\widehat{p}^2_2 = p^2_{{\rm W}^+}$.  The
last condition ensures that the off-shell invariant of the ${\rm W}^+$
boson is left untouched (as the top-quark invariants in the on-shell
projection with two W bosons explained above).  The projection of the
antibottom quark and ${\PW}^-$~boson can be constructed in the same
way.  The decay products of the $\PW^+$ boson (in a similar way to
what has been  done  for the previous on-shell projection) read:
\eq{\widehat{p}_{{\rm e}^+} = p_{{\rm e}^+} \frac{p^2_{{\rm W}^+}}{2 \widehat{p}_{{\rm W}^+} p_{{\rm e}^+}} , \qquad \widehat{p}_{\nu_e} = \widehat{p}_{{\rm W}^+} -\widehat{p}_{{\rm e}^+} . }
The decay products of the ${\rm W}^-$ boson can be handled in the same way.

Concerning the non-factorisable corrections, the notations differ
slightly from the case considered in
Eqs.~\eqref{eq:ensemblesNFC}--\eqref{eq:deltanf}.  In particular, the
ensembles of initial-state, decay-product, and remaining final-state
particles are:
\eq{ I = \{1,2\}, \qquad R_1=\{3,4,5\}, \qquad R_2=\{6,7,8\}, \qquad \text{and} \qquad N= \text{\O} .}
The convention for the sign factors and charges is as in
 Eqs.~ \eqref{eq:sigmaNFC} and \eqref{eq:chargesNFC}.  The expression for
$\delta_{\mathrm{nfact}}$ is still the same as in Eq.~(\ref{eq:deltanf}),
only the content of the ensembles $R_i$ and $N \cup I$ is modified.

Concerning the interference contributions, as for the case of the WW
DPA, the factorisable corrections and the $I$-operator terms have to
be computed to in the pole approximation.  Here, non-factorisable
corrections appear as there are QCD corrections linking the production
part and decay part of the top quarks.  These non-factorisable QCD
corrections can be computed in the same manner as the EW ones.  To do
this, one replaces the charges and matrix elements squared by the
colour-correlated matrix elements squared in Eq.~\eqref{eq:deltanf}.
The non-factorisable QCD contribution thus reads:
\begin{align}
        2 \mathrm{Re} \biggl\{ \mathcal{M}^*_{\mathrm{LO,PA}} & \mathcal{M}_{\mathrm{virt,nfact,PA}}^{\mathrm{QCD}} \biggr\}  =
        \begin{aligned}[t]
                &  - \sum_{a \in R_1} \sum_{b \in R_2} \overline{\mathcal{A}^2_\mathrm{c}} \left(a,b\right) Q^{\mathrm{c}}_a Q^{\mathrm{c}}_b \frac{\alpha}{\pi} \mathrm{Re} \left\{ \Delta\left(i=1, a; j=2, b\right) \right\} \nonumber
        \end{aligned} \\
        &\quad{} -
        \begin{aligned}[t]
                & \sum_{i = 1}^2 \sum_{a \in R_i} \sum_{b \in N \cup I} \overline{\mathcal{A}^2_{\mathrm{c}}} \left(a,b\right) Q^{\mathrm{c}}_a Q^{\mathrm{c}}_b \frac{\alpha}{\pi} \mathrm{Re} \left\{ \Delta_{\mathrm{xf}} \left(i, a; b\right) + \Delta_{\mathrm{xm}}\left(i; b\right) \right\} , \\
        \end{aligned}
\end{align}
where $\overline{\mathcal{A}^2_{\mathrm{c}}} \left(a,b\right)$ denotes
the colour-correlated squared amplitude between particle $a$ and $b$
as defined in \citere{Actis:2016mpe}.  The charges
$Q^{\mathrm{c}}_{a/b}$ take the value 1 or 0 if the particle carries a
colour charge or not, respectively.

\subsection{Validation}
\label{sec:Validations}

\begin{sloppy}
  Several checks have been performed on this computation.  All
  tree-level, \ie Born and real, matrix elements squared have been
  compared with the code \madgraph~\cite{Alwall:2014hca}.  Out of 4000
  phase-space points, more than 99.9 $\%$ agree to 11 and 10 digits
  for the Born and real matrix elements squared, respectively.  All
  hadronic Born cross sections ($\Pg \Pg$, $\bar{q}q$ and $\Pg \gamma$ channels)
  have been compared with \madgraph, and agreement within the
  integration error has been found.
\end{sloppy}

IR und ultra-violet (UV) finiteness have been verified by calculating
the cross section for different IR and UV regulators, respectively.
The implementation of the dipole subtraction method has been checked
by varying the $\alpha$ parameter\footnote{For the results presented
  in this paper the value $\alpha= 10^{-2}$ has been used.} from
$10^{-2}$ to 1.  The parameter $\alpha$ allows one to improve the
numerical stability of the integration by restricting the phase space
for the dipole subtraction terms to the vicinity of the singular
regions \cite{Nagy:1998bb}.

\begin{figure}
\begin{center}
                        \includegraphics[width=0.7\linewidth]{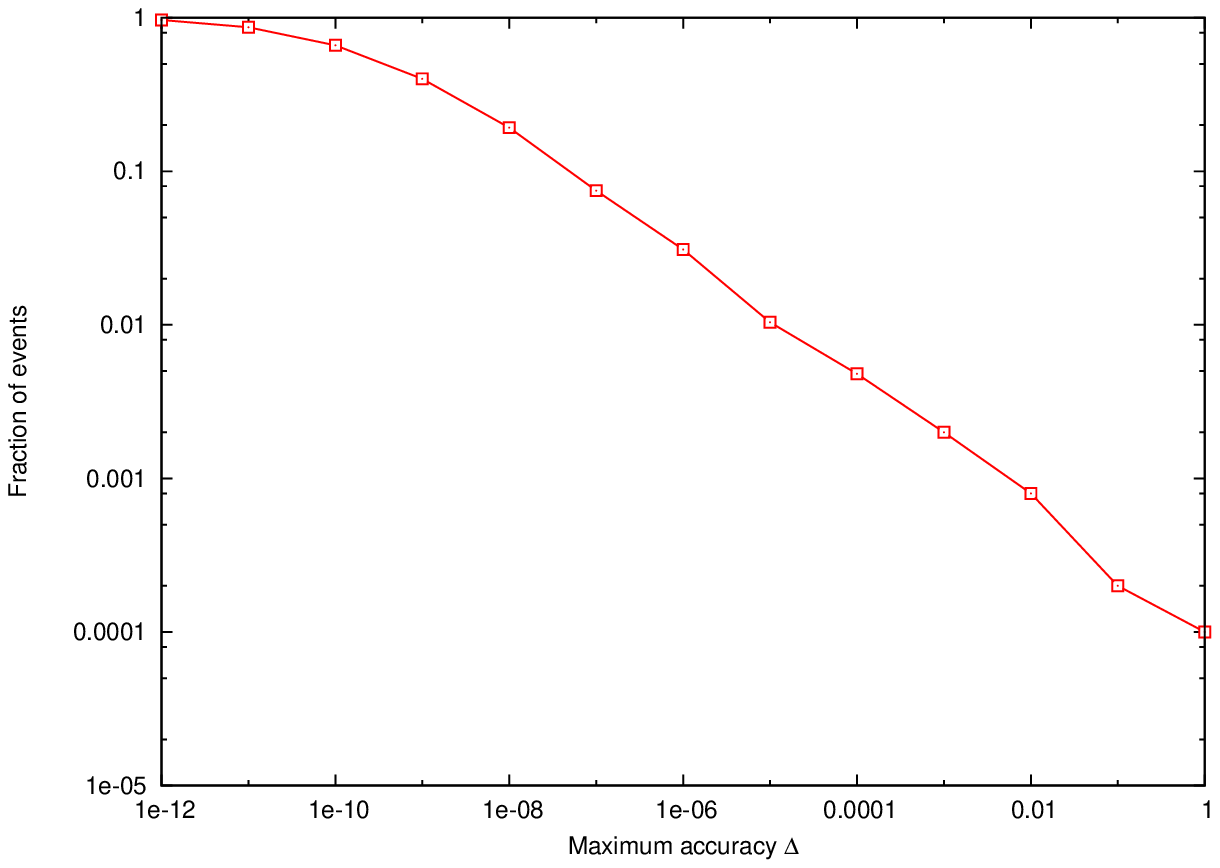}
\end{center}
        \caption{\label{Ward}%
                Cumulative fraction of events with a relative accuracy larger than $\Delta$ for $\Pp\Pp\to\Pe^+\nu_\Pe \mu^-\bar{\nu}_\mu\Pb\bar{\Pb}$ at NLO EW.
        }
\end{figure}

The virtual corrections have been scrutinised in several ways.  First,
the computer code \recola allows for an internal check of a Ward
identity. One can substitute the polarisation vector of one of the
initial-state gluons by its momentum normalised to its energy, \ie
$\epsilon^\mu_\Pg \to p^\mu_\Pg/p^0_\Pg$, in the one-loop amplitude.
The cumulative fraction of events with
$\Re\mathcal{M}^*_0(\epsilon_\Pg)\mathcal{M}_1(\epsilon_\Pg\to
p_\Pg/p^0_\Pg)/\Re\mathcal{M}^*_0(\epsilon_\Pg)\mathcal{M}_1(\epsilon_\Pg)>\Delta$
is plotted in Fig.~\ref{Ward}.  It gives results comparably good to
those of \citere{Denner:2015yca} for $\Pp\Pp\to\Pe^+\nu_\Pe
\mu^-\bar{\nu}_\mu\Pb\bar{\Pb} \PH$ where the median is also around
$10^{-9}$.  Second, thanks to the two libraries implemented in the
computer code \collier, we have been able to estimate the potential
error induced when evaluating the virtual corrections.  This turned
out to be below the per-mille level after integration, \ie below the
precision of integration we have required for the numerical results.
Finally, the excellent agreement found with one of the two DPAs (see
below) for the observables computed confirms that the full one-loop
amplitudes used in this computation are reliable.  Note that we have
checked also our implementation of the (double-)pole approximation for
a variety of processes ranging from Drell--Yan (with W and Z boson) to
di-boson production (also involving W or Z bosons).

\section{Numerical Results}
\label{sec:results}

\subsection{Input parameters and selection cuts}
\label{ssec:InputParameters}

In this section, integrated cross sections and differential
distributions including NLO EW corrections for the LHC at a
centre-of-mass energy $\sqrt{s}=13\TeV$ are presented.  For the parton
distribution functions, LHAPDF
6.1.5~\cite{Andersen:2014efa,Buckley:2014ana} has been employed.
Specifically, the
$\mathrm{NNPDF23}\_\mathrm{nlo}\_\mathrm{as}\_0119\_\mathrm{qed}$
set~\cite{Ball:2013hta,Carrazza:2013bra,Carrazza:2013wua} at NLO QCD
and LO QED has been used for all the LO and NLO contributions.  This
features the inclusion of a photon PDF needed for the photon-initiated
contributions.  The strong coupling constant $\alphas$ is provided by
the PDF set based on a two-loop QCD running with a dynamical flavour
scheme with $N_\text{F}=6$ active flavours.\footnote{Note that the
  difference between the fixed 5-flavour scheme and the dynamical
  6-flavour scheme can reach a few per cent above the top-mass
  threshold \cite{Demartin:2010er}.}  For the fixed renormalisation
and factorisation scale $\mu_\text{fix}=m_{\rm t}$, we find
\begin{equation}\label{eqn:Alphas}
    \alphas(\mu_\text{fix}) = 0.1084656\ldots\;.
\end{equation}
Note that contributions for bottom-quark PDFs have been neglected.

Concerning the electromagnetic coupling $\alpha$, the $G_\mu$ scheme
\cite{Denner:2000bj} has been used where $\alpha$ is obtained from the
Fermi constant,
\begin{equation}\label{eqn:FermiConstant}
  \alpha = \frac{\sqrt{2}}{\pi} G_\mu \MW^2 \left( 1 - \frac{\MW^2}{\MZ^2} \right)
  \qquad \text{with}  \qquad   \GF    = 1.16637\times 10^{-5}\GeV.            
\end{equation}

The input parameters are taken from \citere{Beringer:1900zz}, and the
numerical values for the masses and widths used in this computation
read:
\begin{alignat}{2} \label{eqn:ParticleMassesAndWidths}
                 \Mt   &=  173.34\GeV,       & \quad \quad \quad \Gt &= 1.36918\ldots\GeV,  \nonumber \\
                \MZOS &=  91.1876\GeV,      & \GZOS &= 2.4952\GeV,  \nonumber \\
                \MWOS &=  80.385\GeV,       & \GWOS &= 2.085\GeV,  \nonumber \\
                M_{\rm H} &=  125.9\GeV.       & 
\end{alignat}
The masses and widths of all other quarks and leptons have been
neglected.  We have verified that the effect of a finite bottom-quark
mass on the cross section is below the per-cent level in our set-up.
The top-quark width has been taken from \citere{Basso:2015gca}, where
it has been calculated including both EW and QCD NLO corrections for
massive bottom quarks. We have found that the effect of the
bottom-quark mass on the top-quark width is at the per-mille level by
computing the leptonic partial decay width of the top-quark using
\citere{Jezabek:1988iv} with massive and massless bottom quarks.  Such
differences are irrelevant with respect to the integration errors for
the cross section.
We have chosen to use the same top width for our calculation at LO and NLO, since this allows to improve 
QCD calculations upon multiplying with our results for the relative EW correction factors.

The measured on-shell (OS) values for the masses and widths of the W
and Z bosons are converted into pole values for the gauge bosons
($V=\PW,\PZ$) according to \citere{Bardin:1988xt},
\newcommand{\MVOS}{\ensuremath{M_V^\text{OS}}\xspace}%
\newcommand{\GVOS}{\ensuremath{\Gamma_V^\text{OS}}\xspace}%
\begin{equation}
        M_V = \MVOS/\sqrt{1+(\GVOS/\MVOS)^2}\,,\qquad  \Gamma_V = \GVOS/\sqrt{1+(\GVOS/\MVOS)^2}.
\end{equation}

The QCD jets are clustered using the anti-$k_\text{T}$ algorithm
\cite{Cacciari:2008gp}, which is also used to cluster the photons with
light charged particles, with jet-resolution parameter $R=0.4$.  The
distance between two particles $i$ and $j$ in the rapidity--azimuthal-angle
plane is defined as
\begin{equation}\label{eqn:DeltaR}
        R_{ij} = \sqrt{(\Delta \phi_{ij})^2+(y_i-y_j)^2},
\end{equation}
where $\Delta \phi_{ij}$ is the azimuthal-angle difference. The
rapidity of jet $i$ is given by $y_i=\frac{1}{2}\ln
\frac{E+p_z}{E-p_z}$ with the energy $E$ of the jet and the component
of its momentum along the beam axis $p_z$.  Only final-state quarks,
gluons, and charged fermions with rapidity $|y|<5$ are clustered into
IR-safe objects.

After recombination, standard selection cuts on the transverse momenta
and rapidities of charged leptons and b~jets, missing transverse
momentum and rapidity--azimuthal-angle distance between b~jets
according to Eq.~\eqref{eqn:DeltaR} are imposed.  In the final state,
two b jets\footnote{Bottom quarks in jets lead to bottom jets.} and
two charged leptons are required, and the following selection cuts are
applied:
\begin{alignat}{5} \label{eqn:cuts}
                 \text{b jets:}                     && \qquad \ptsub{\Pb}         &>  25\GeV,  & \qquad |y_\Pb|   &< 2.5, & \nonumber \\
                \text{charged lepton:}              && \ptsub{\Pl}         &>  20\GeV,  & |y_{\Pl}| &< 2.5, &\nonumber \\
                \text{missing transverse momentum:} && \ptsub{\text{miss}} &>  20\GeV,                      &\nonumber \\
                \text{b-jet--b-jet distance:}       && \Delta R_{\Pb\Pb}   &> 0.4.                          &
\end{alignat}

\subsection{Integrated cross section}
\label{ssec:IntegratedCrossSection}

In this section the results for the integrated cross section are
discussed.  The different contributions are summarised in
\refta{table:results_summary} for the LHC running at a centre-of-mass
energy of $\sqrt{s}=13\TeV$.  It corresponds to the input parameters
given in Eqs.~\eqref{eqn:Alphas}--\eqref{eqn:ParticleMassesAndWidths}
, while the selection cuts for this set-up are defined in
Eq.~\eqref{eqn:cuts}.  As stated before, the LO contributions are of
the order $\order{\alphas^2\alpha^{4}}$, while the EW NLO corrections
are of the order $\order{\alphas^2\alpha^{5}}$.  Since the ${\rm
  \gamma g}$ contribution is of the order $\order{\alphas
  \alpha^{5}}$, we have not included them in the definition of the EW
NLO corrections.  Nonetheless we give it for reference.
\begin{table}
\begin{center} 
\begin{tabular}{ c  c  c   c }
 Ch. & $\sigma_{\rm LO}$ [fb] & $\sigma_{\rm NLO \; EW}$ [fb] & $\delta$ [$ \% $]\\
  \hline\hline
$\Pg \Pg$        &  $2824.2(2)$ & $2834.2(3) $ & 0.35 \\
${ q \bar{q}}$      &  $375.29(1) $ & $377.18(6) $ & 0.50 \\
${\rm g} q(/\bar{q})$    &                                   & $0.259(4)  $ &          \\
  \hline
${\rm \gamma g}$  &                                   & $ 27.930(1)    $ &     \\
\hline\hline
$\Pp\Pp$         &  $ 3199.5(2) $ & $3211.7(3) $ & $0.38$ \\
  \hline
\end{tabular}
\end{center}
        \caption[Composition of the integrated cross section]{\label{table:results_summary}
                Different contributions to the integrated cross section for $\Pp\Pp \to 
                \Pe^+\nu_\Pe \mu^- \bar{\nu}_\mu \Pb \bar{\Pb} (\Pj)$ at a centre-of-mass energy of $\sqrt{s}=13\TeV$.
                The quark--antiquark contributions comprise $ q=\Pu,\Pd,\Pc,\Ps$.
                The channel ${\rm g} q(/\bar{q})$ denotes the real
                radiation of a quark or an antiquark.
                In the total cross section (denoted by $\Pp\Pp$), the photon-induced channel (denoted by ${\rm \gamma g}$) has not been included.
                The relative correction is defined as $\delta =
                \sigma_{\rm NLO \; EW}/\sigma_{\rm LO}$.
                Integration errors of the last digits are given in parentheses.}
\end{table}

At the LHC (in contrast to the Tevatron) the gluon--gluon-initiated
channel is dominant owing to the enhanced gluon PDF.  The ${ q
  \bar{q}}$ channels that comprise ${ q}=\Pu,\Pd,\Pc,\Ps$ are one
order of magnitude smaller and represent only $11.7 \%$ of the total
integrated cross section (both at LO and NLO).  The corrections to
these two channels are $0.35 \%$ and $0.50 \%$, respectively.
Moreover, the ${\rm g} q/\bar{q}$ channel contributes only at the
sub-per-mille level, being of the order of the error on the integrated
cross section.  The EW corrections to the full partonic process amount
to $0.38 \%$.

For on-shell top-pair production the EW corrections are usually
between $-1 \%$ and $-2 \%$ (see \citere{Pagani:2016caq} for a recent
evaluation).  This difference to our results can be explained by the
EW corrections to the top-quark width that are implicitly contained in
our calculation and amount to 1.3\% \cite{Basso:2015gca}. Since we use
the same value for the width in the resonant top-quark propagators at
LO and NLO, this effect does not cancel.  Subtracting twice the
relative NLO corrections to the top width from our corrections yields
a correction to top-pair production of the usual size.

The ${\rm \gamma g}$ channel gives a contribution of the
order of one per cent.  Thus, calculating QCD corrections to
this partonic channel would lead at most to a per-mille contribution.
Nonetheless, the photon-induced channel represents a non-negligible
contribution to the cross section.

As stated before we have considered massless bottom quarks and have
neglected their PDF contributions.  To justify this, we have computed
the LO hadronic cross sections including massive bottom quarks and
bottom-quark PDFs.  The effect of a finite bottom-quark mass is at the
level of $0.8 \%$.  The bottom PDFs contribute at the level of $0.01
\%$ to the process $\Pp\Pp\to\Pe^+\nu_\Pe
\mu^-\bar{\nu}_\mu\Pb\bar{\Pb}$ at LO. This tiny contribution is
explained by the dominance of the gluon PDFs.

Thus, the EW corrections are below the per-cent level for the
integrated cross section.  However, as shown in the next section, this
statement does not hold for differential distributions.

\subsection{Differential distributions}
\label{ssec:DifferentialDistributions}

Turning to differential distributions, we show two plots for each
observable.  
The upper panels display the LO and NLO EW predictions, while the
lower panels show the relative correction
$\delta = \sigma_\text{NLO EW} /
\sigma_\text{LO} - 1$ in per cent.  In addition the ${\rm \gamma g}$
contribution is depicted as $\delta_{\rm \gamma g} = \sigma_{\rm
  \gamma g} / \sigma_\text{LO}$ and labelled by {\em{photon}}.

\begin{figure}
        \setlength{\parskip}{-10pt}
        \begin{subfigure}{0.49\textwidth}
                \subcaption{}
                \includegraphics[width=\textwidth]{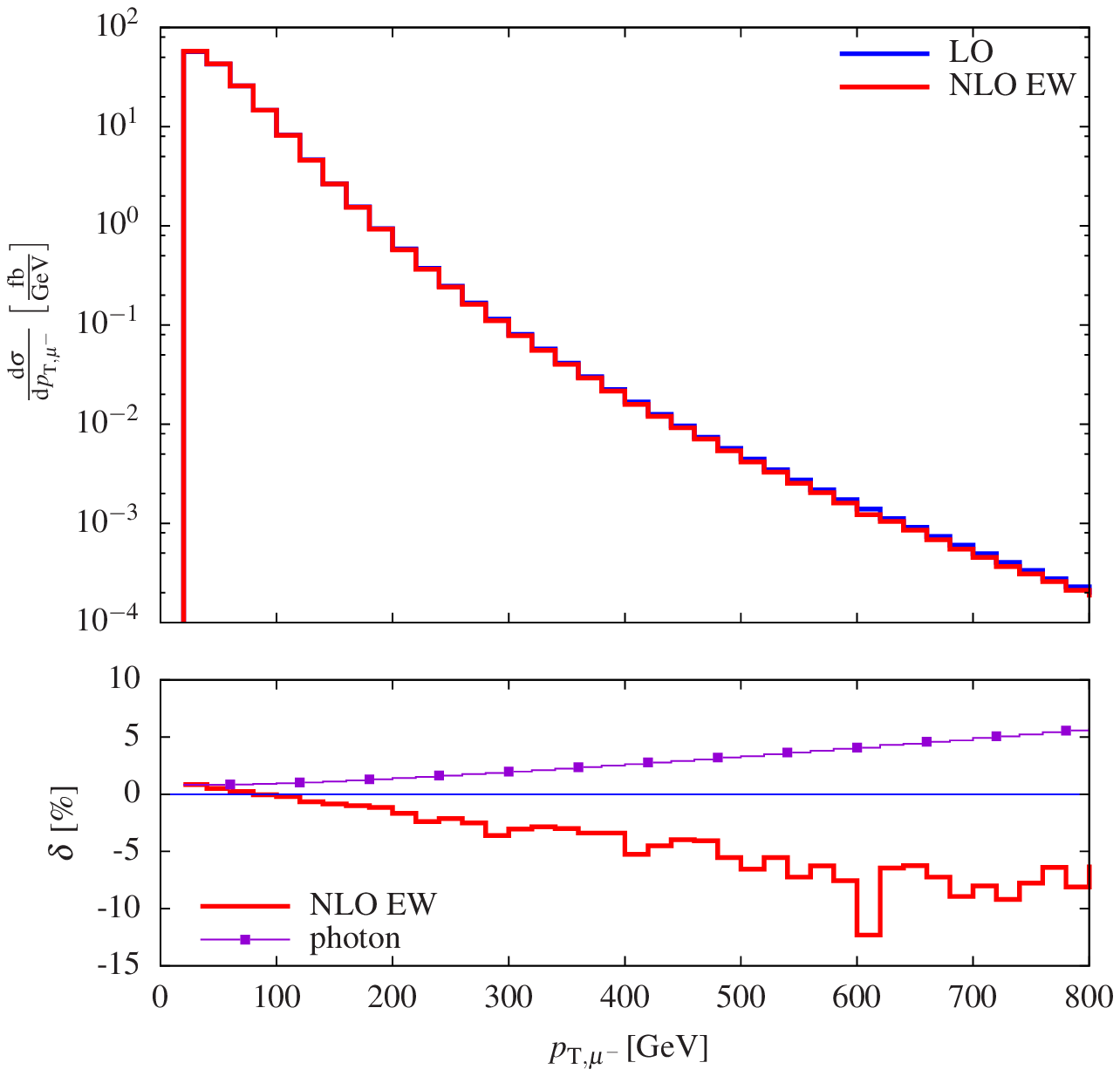}
                \label{plot:transverse_momentum_positron}
        \end{subfigure}
        \hfill
        \begin{subfigure}{0.49\textwidth}
                \subcaption{}
                \includegraphics[width=\textwidth]{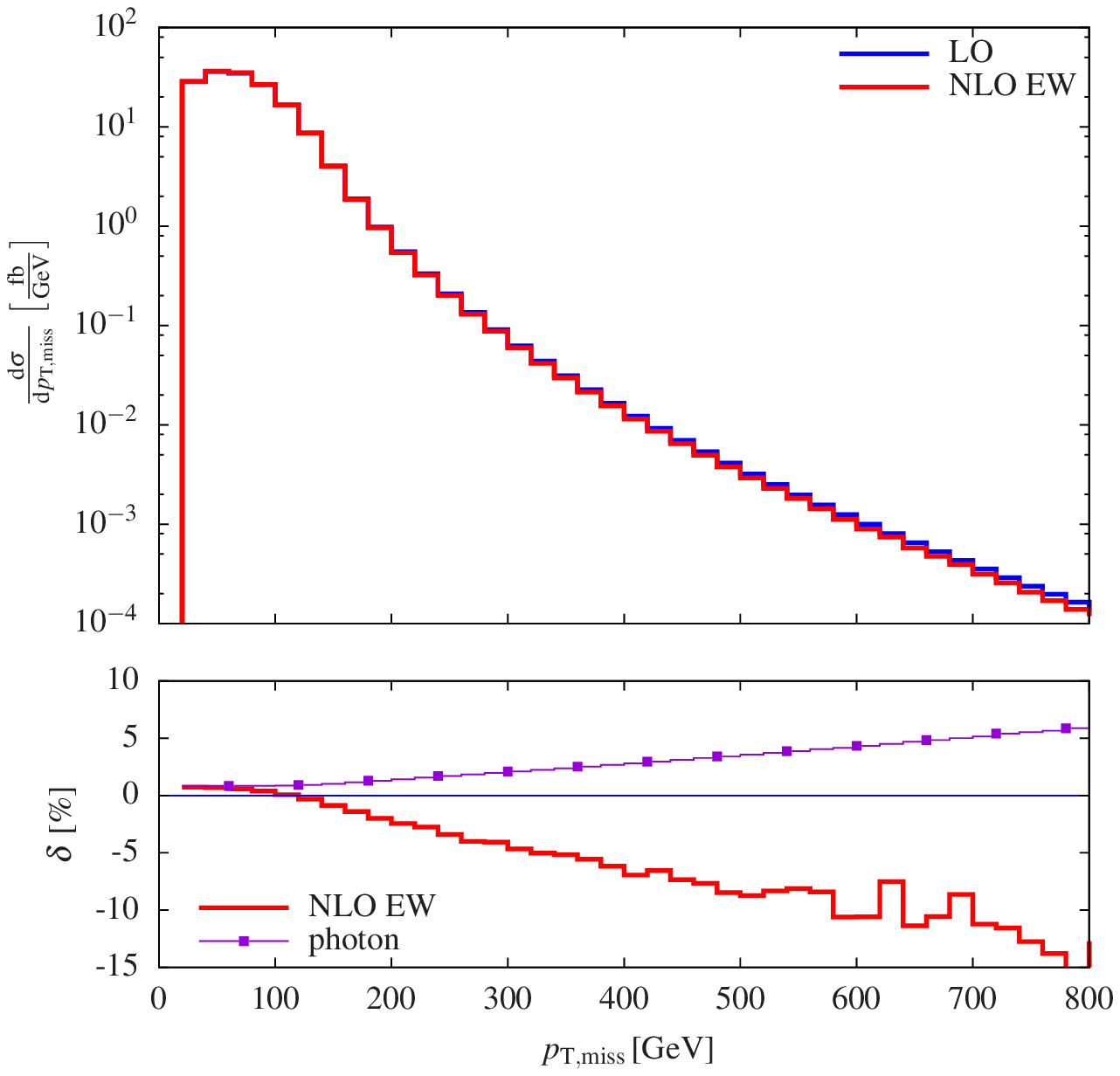}
                \label{plot:transverse_momentum_truth_missing} 
        \end{subfigure}
        
        \begin{subfigure}{0.49\textwidth}
                \subcaption{}
                \includegraphics[width=\textwidth]{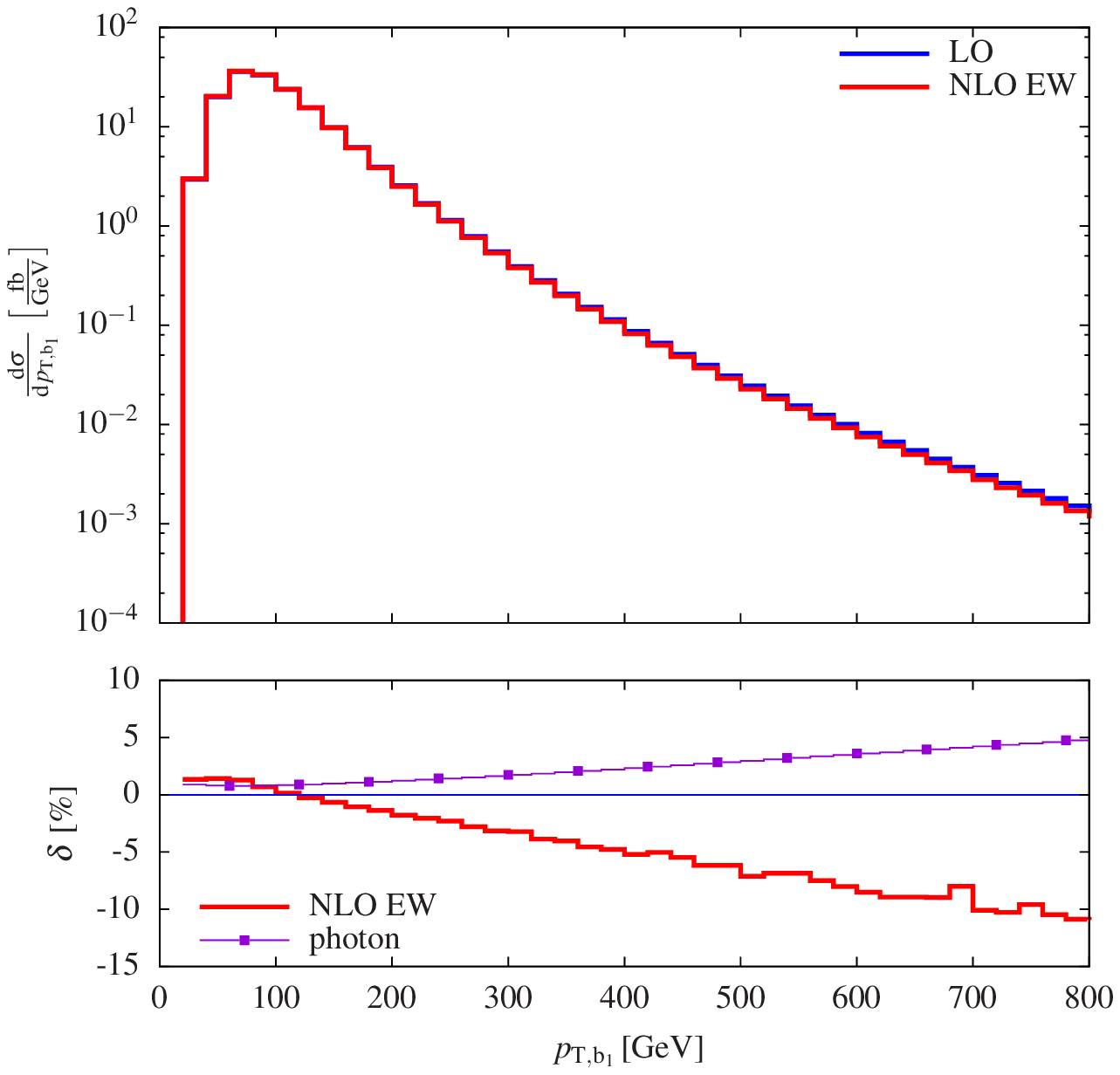}
                \label{plot:transverse_momentum_b1}
        \end{subfigure}
        \hfill
        \begin{subfigure}{0.49\textwidth}
                \subcaption{}
                \includegraphics[width=\textwidth]{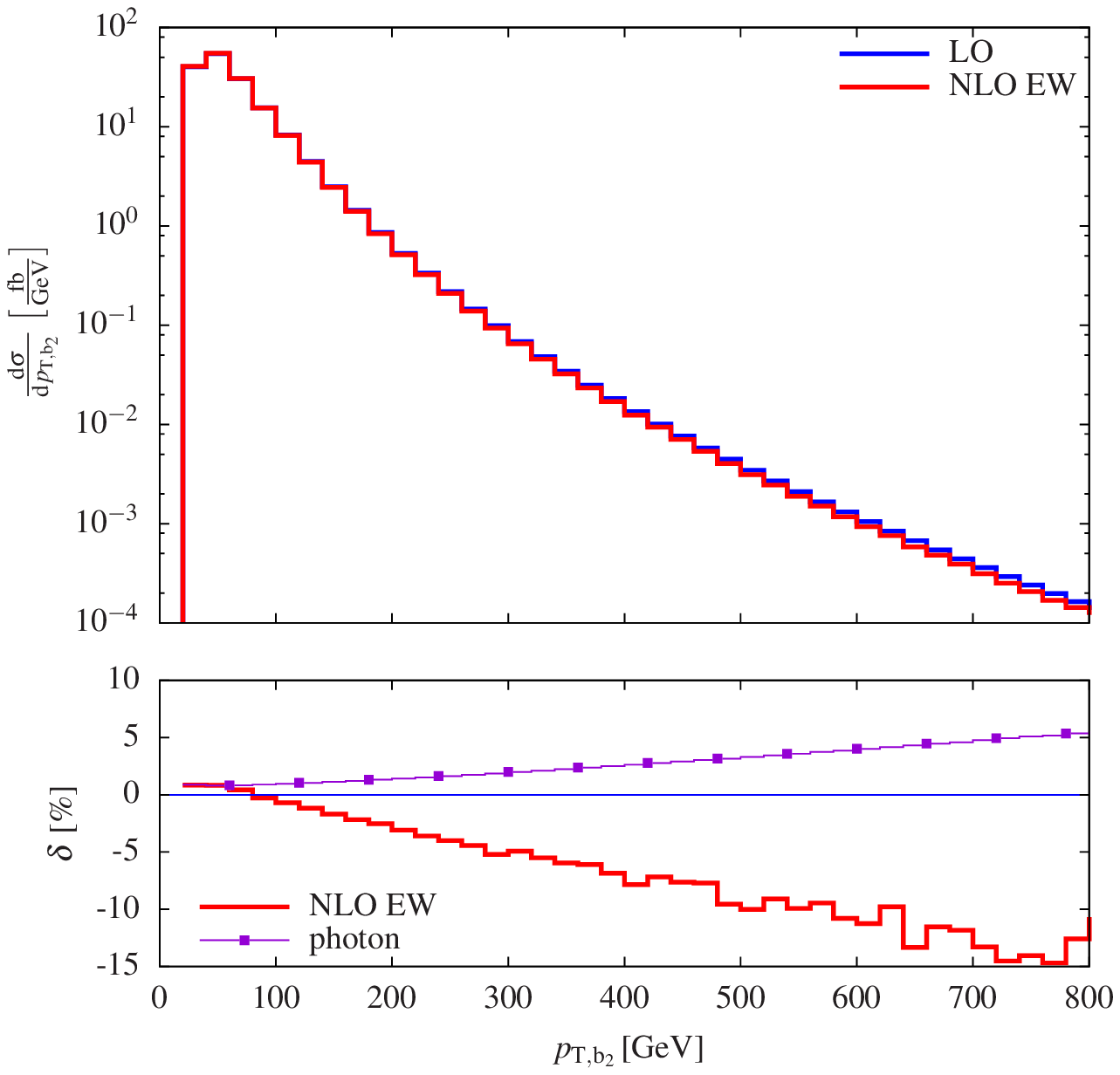}
                \label{plot:transverse_momentum_b2}
        \end{subfigure}
        
        \begin{subfigure}{0.49\textwidth}
                \subcaption{}
                \includegraphics[width=\textwidth]{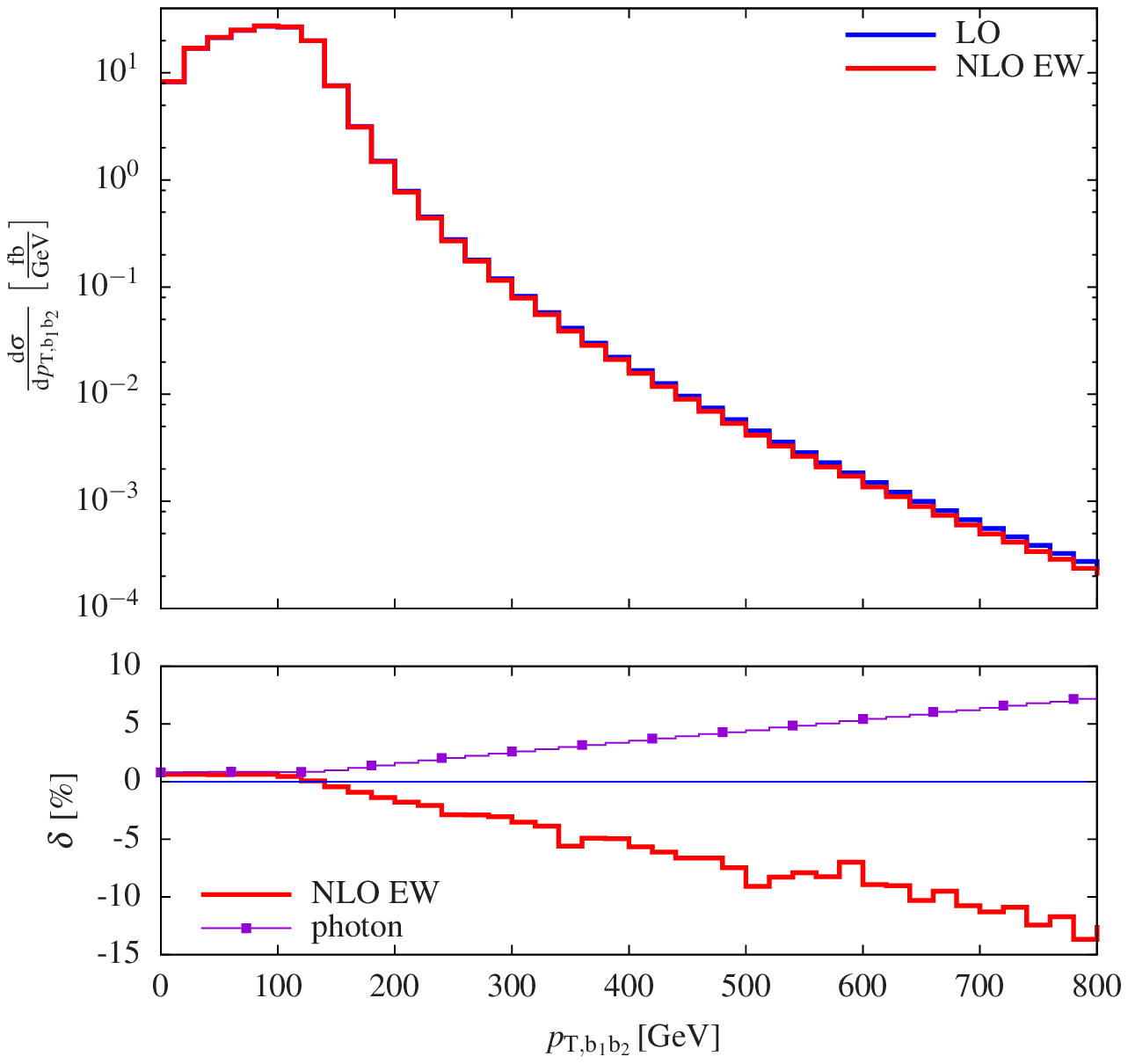}
                \label{plot:transverse_momentum_bb12}
        \end{subfigure}
        \hfill
        \begin{subfigure}{0.49\textwidth}
                \subcaption{}
                \includegraphics[width=\textwidth]{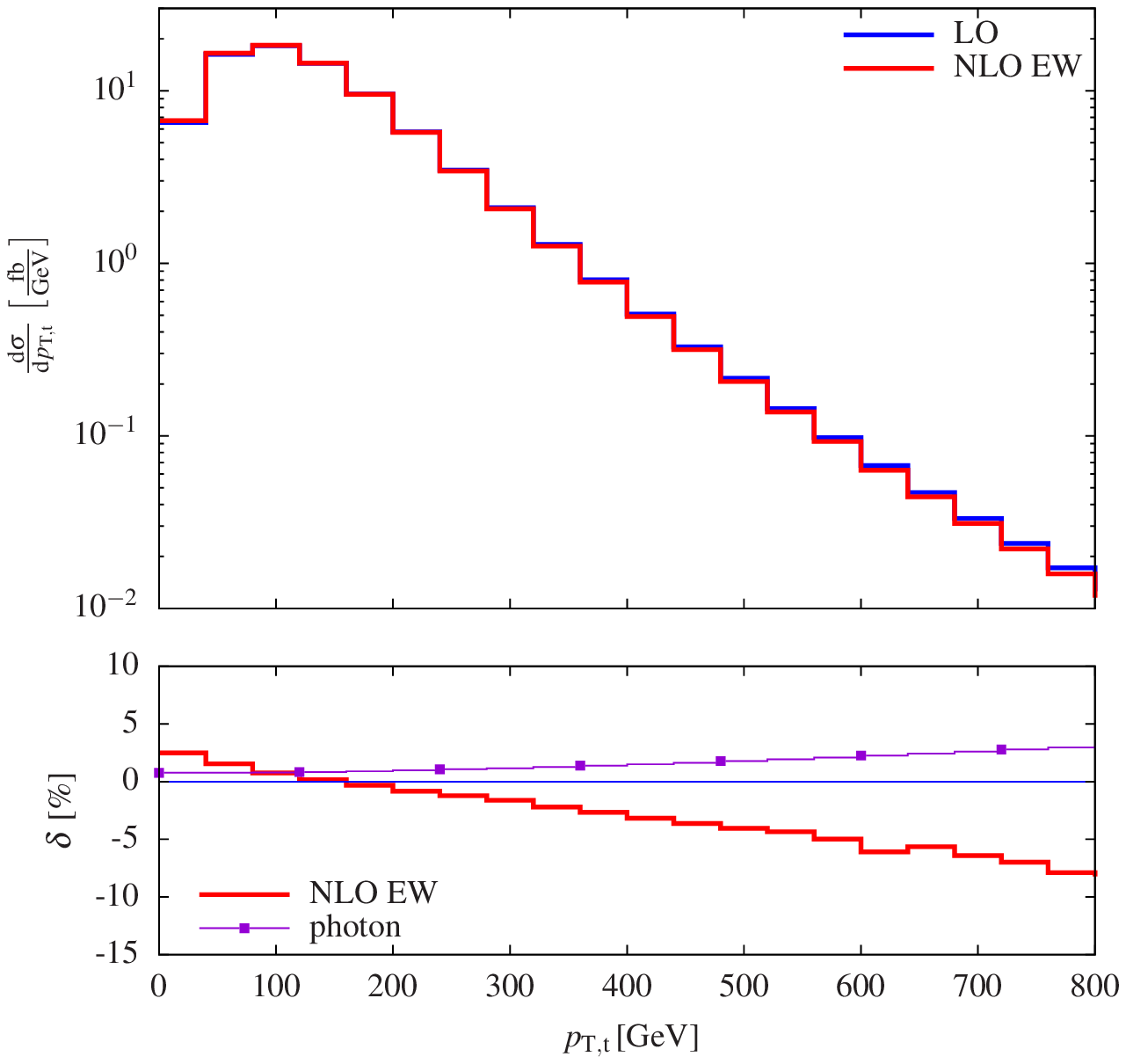}
                \label{plot:transverse_momentum_top}
        \end{subfigure}
     
        \vspace*{-3ex}
        \caption{\label{fig:transverse_momentum_distributions}%
                Transverse-momentum distributions at a centre-of-mass energy $\sqrt{s}=13\TeV$ at the LHC: 
                \subref{plot:transverse_momentum_positron} for the muon~(upper left), %
                \subref{plot:transverse_momentum_truth_missing} for missing momentum~(upper right), %
                \subref{plot:transverse_momentum_b1} for the harder \Pb~jet~(middle left), %
                \subref{plot:transverse_momentum_b2} for the softer \Pb~jet~(middle right), %
                \subref{plot:transverse_momentum_bb12} for the \Pb-jet pair~(lower left), and %
                \subref{plot:transverse_momentum_top} for the reconstructed top quark~(lower right).
                The lower panel shows the relative NLO EW correction
                $\delta = \sigma_{\text{NLO EW}} / \sigma_\text{LO} -
                1$ and the relative photon-induced
                contributions $\delta = \sigma_{ \gamma g } /
                \sigma_\text{LO} $ in per cent.  }
\end{figure}%
Figure~\ref{plot:transverse_momentum_positron} displays the
distribution of the muon transverse momentum, while
\reffis{plot:transverse_momentum_b1} and
\ref{plot:transverse_momentum_b2} show the transverse momenta of the
harder and softer bottom quark (according to $\pt$ ordering).  In
\reffi{plot:transverse_momentum_truth_missing} we present the
distribution in the missing transverse momentum, defined as the sum of
the transverse momenta of the two neutrinos, {\emph i.e.}
$\ptsub{\text{miss}}=\abs{\vec{p}_{\text{T},\nu_{\Pe}}+\vec{p}_{\text{T},\bar{\nu}_{\mu}}}$.
The transverse momentum of the bottom-jet pair is displayed in
\reffi{plot:transverse_momentum_bb12} and the one of the reconstructed
top quark in \reffi{plot:transverse_momentum_top}.  In all
distributions in \reffi{fig:transverse_momentum_distributions} one can
clearly see the effects of the Sudakov logarithms at high transverse
momenta.  In general, the corrections are within $2 \%$ for transverse
momenta below 50\GeV and grow negative towards high transverse
momenta. The EW corrections account for effects of up to $15\%$ over
the considered phase-space range up to $800\GeV$.  In all
transverse-momentum distributions, the gluon--photon-induced channel
increases towards the high-momentum region.  This is due to the fact
that the photon PDF grows faster than the quark and gluon PDFs in this
region \cite{Pagani:2016caq}.  Indeed, the photon-induced
contributions typically reach $5$--$6\%$ at $\pt=800\GeV$.
But as the photon PDF is still poorly known
\cite{Carrazza:2013bra,Ball:2013hta}, this statement should be
understood with caution.  
More specifically, in the transverse-momentum distribution of the
softer bottom quark, the EW corrections go from $2 \%$ at low
transverse momentum down to $- 15 \%$ at 800\GeV.  There, the
photon-induced channel accounts for $1 \%$ at low transverse momentum
and up to $5 \%$ at 800\GeV.

\begin{figure}
        \setlength{\parskip}{-10pt}
        \begin{subfigure}{0.49\textwidth}
                \subcaption{}
                \includegraphics[width=\textwidth]{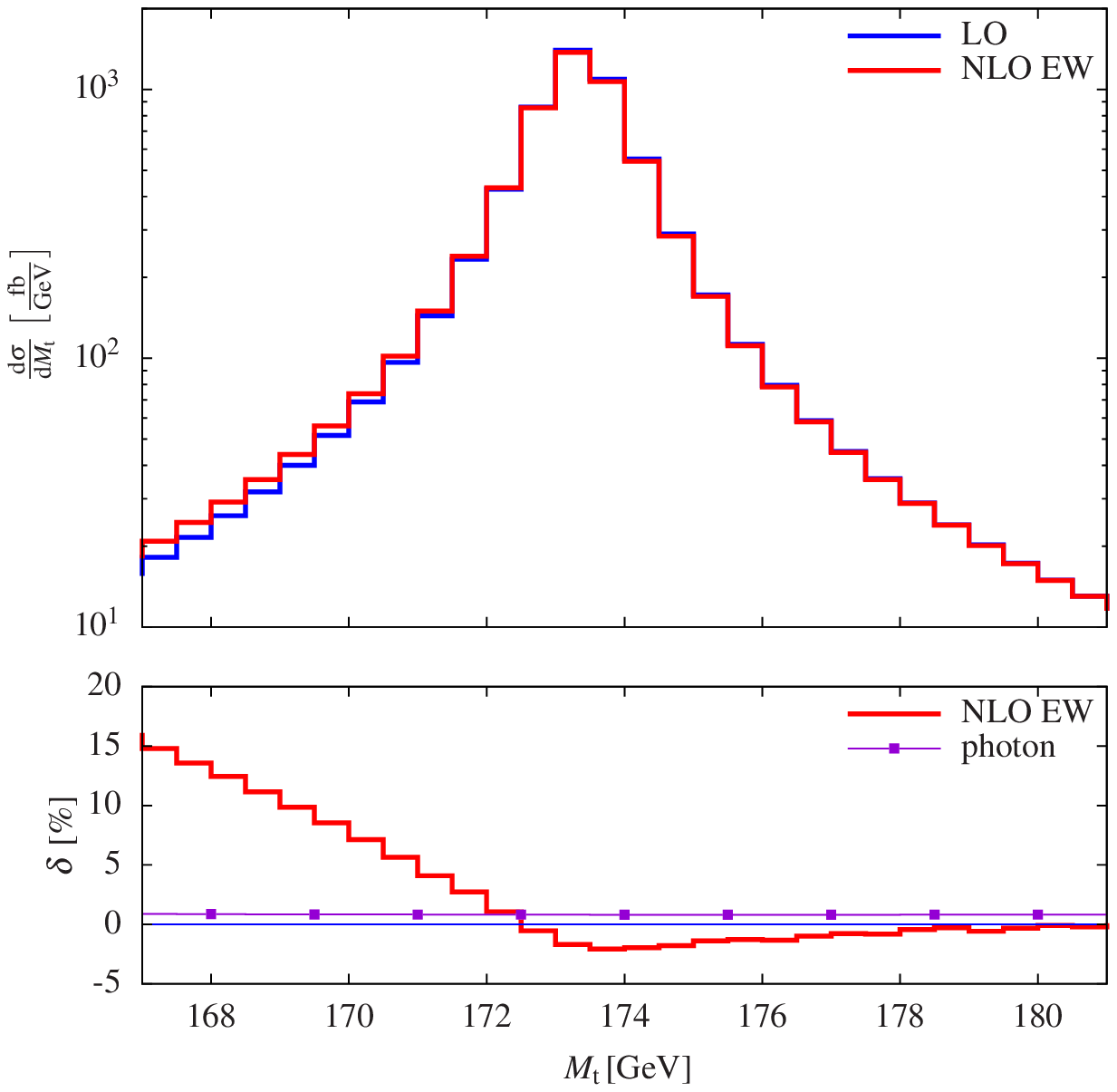}
                \label{plot:invariant_mass_truth_top}
        \end{subfigure}
        \hfill
        \begin{subfigure}{0.49\textwidth}
                \subcaption{}
                \includegraphics[width=\textwidth]{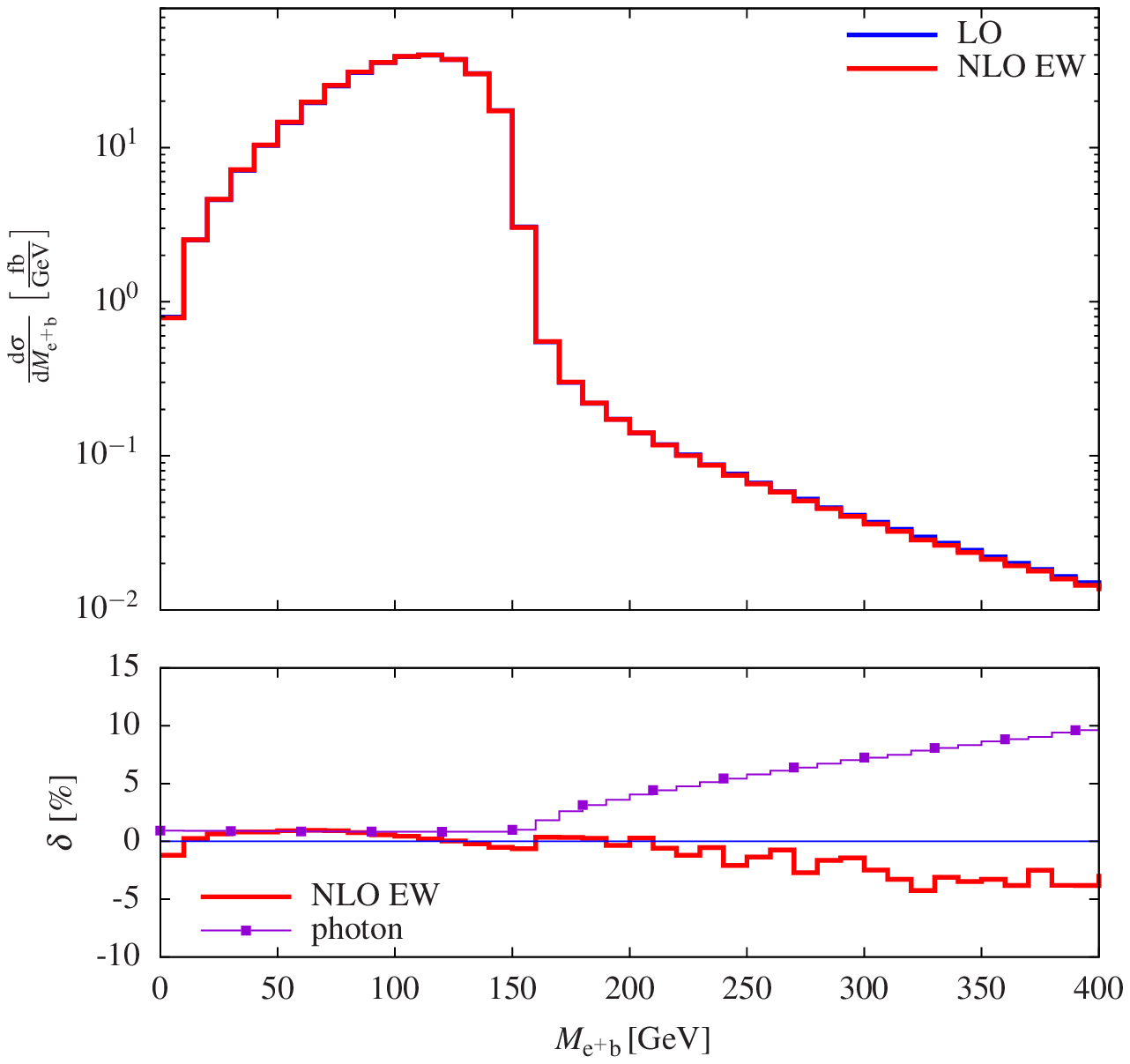}
                \label{plot:invariant_mass_truth_epb}
        \end{subfigure}
        \begin{subfigure}{0.49\textwidth}
                \subcaption{}
                \includegraphics[width=\textwidth]{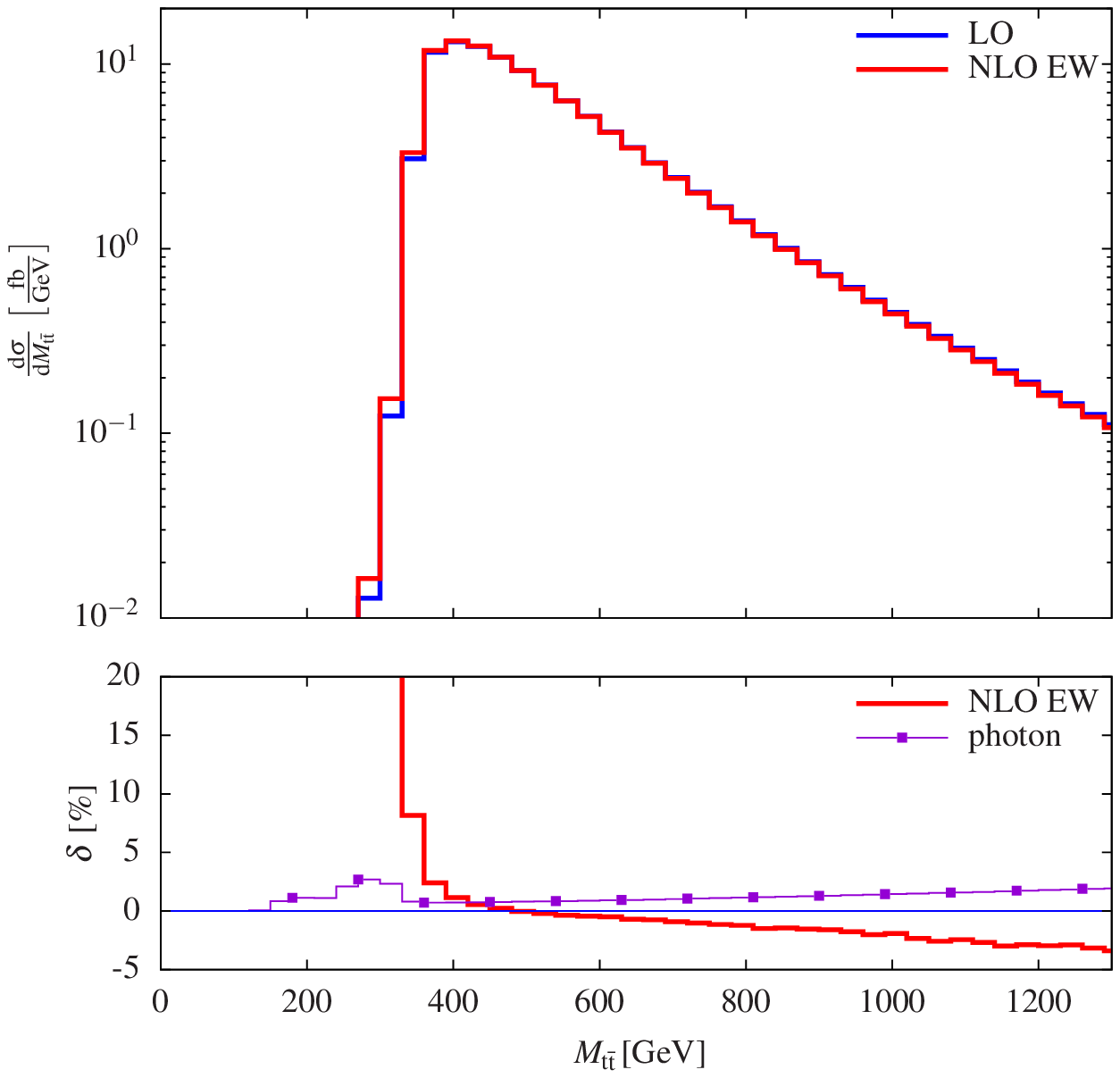}
                \label{plot:invariant_mass_truth_ttx}
        \end{subfigure}
        \hfill
        \begin{subfigure}{0.49\textwidth}
                \subcaption{}
                \includegraphics[width=\textwidth]{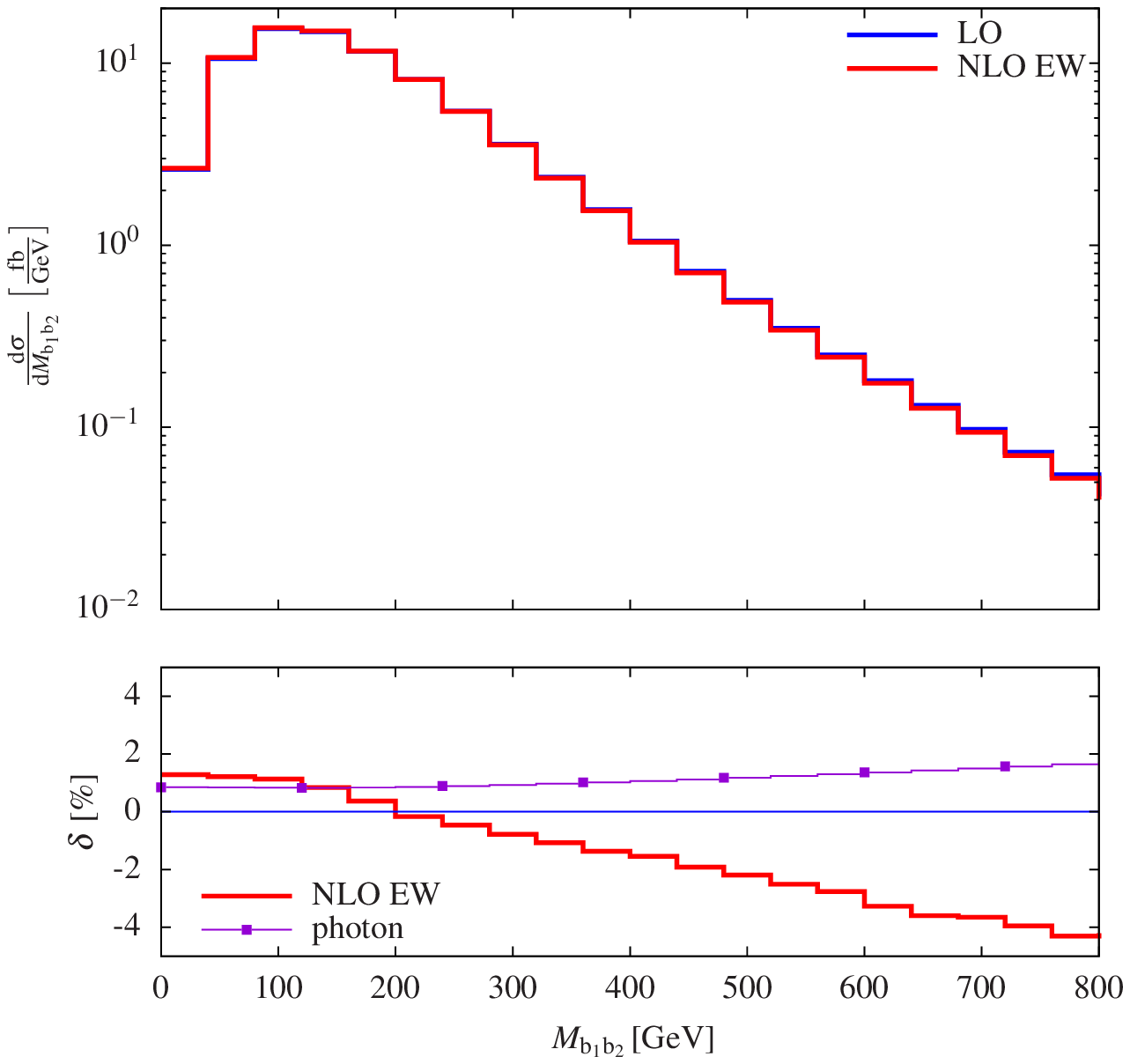}
                \label{plot:invariant_mass_bb12} 
        \end{subfigure}
        
        \vspace*{-3ex}
        \caption{\label{fig:invariant_mass_differential_distributions}%
                Differential distributions at a centre-of-mass energy $\sqrt{s}=13\TeV$ at the LHC: 
                 \subref{plot:invariant_mass_truth_top}~invariant mass of the reconstructed top quark~(upper left),
                 \subref{plot:invariant_mass_truth_epb}~invariant mass of the $\Pe^+ \Pb$ system (upper right),
                \subref{plot:invariant_mass_truth_ttx}~invariant mass of the reconstructed $\Pt\bar{\Pt}$
                system  (lower left), and %
                \subref{plot:invariant_mass_bb12}~invariant mass of
                the \Pb-jet pair (lower right).  The lower panel shows
                the relative NLO EW correction $\delta =
                \sigma_{\text{NLO EW}} / \sigma_\text{LO} - 1$ and the
                relative photon-induced contributions $\delta =
                \sigma_{ \gamma g } / \sigma_\text{LO} $ in per cent.}
\end{figure}%
In \reffi{fig:invariant_mass_differential_distributions}, a selection
of invariant-mass distributions is shown containing those of the
reconstructed top quark (\reffi{plot:invariant_mass_truth_top}), of
the $\Pe^+ \Pb$ system (\reffi{plot:invariant_mass_truth_epb}), of the
reconstructed $\Pt \bar{\Pt}$ system
(\reffi{plot:invariant_mass_truth_ttx}), and of the $\Pb \bar{\Pb}$
system (\reffi{plot:invariant_mass_bb12}).  Below the top mass, the
corrections to the invariant mass of the reconstructed top quark reach
up to $15 \%$.  Such a radiative tail is also observed in similar
processes at NLO QCD \cite{Denner:2012yc,Denner:2015yca}, and is due
to final-state photons (or gluons) that are not reconstructed with the
decay products of the top quark.  In the distribution in the invariant
mass of the positron--bottom-quark system, which is the invariant mass
of the visible decay products of the top quark, the LO cross section
decreases sharply around $155\GeV$.  This is due to the existence of
an upper bound $M^2_{\Pe^+ \Pb} < M^2_{\rm t} - M^2_{\rm W} \simeq
(154\GeV)^2$ for on-shell top quark and W boson.  This edge is very
sensitive to the top mass and thus allows to determine its
experimental value precisely.  It marks the transition from on-shell
to off-shell top-quark production.  In that regard, higher-order
corrections to this observable are particularly relevant.  At the
threshold near $155\GeV$, the EW corrections are negative and below
one per cent, while the photon-induced contributions reach $1\%$.  The
corrections below this threshold are of the order of $1 \%$.  On the
other hand, above this bound the EW corrections go down to $-4 \%$ for
an invariant mass of 400\GeV, while the photon-induced contributions
grow to $+10 \%$ at $M_{\Pe^+\Pb}=400\GeV$.  Thus, the EW corrections
and photon-induced contributions should be taken into account.  The
invariant mass of the $\Pt \bar{\Pt}$ system is a very important
observable as one could expect new physics in its high-energy tail
\cite{Frederix:2007gi,Backovic:2015soa}.  The corresponding EW
corrections are significant and vary from $1 \%$ at $400\GeV$ to $-4
\%$ at $1300\GeV$.  The invariant mass of the $\Pb \bar{\Pb}$ system
also displays typical EW corrections, accounting for a $5 \%$
variation over the considered range, accompanied by a relatively small
photon-induced contribution below $2 \%$.

\begin{figure}
        \setlength{\parskip}{-10pt}
        \begin{subfigure}{0.49\textwidth}
                \subcaption{}
                \includegraphics[width=\textwidth]{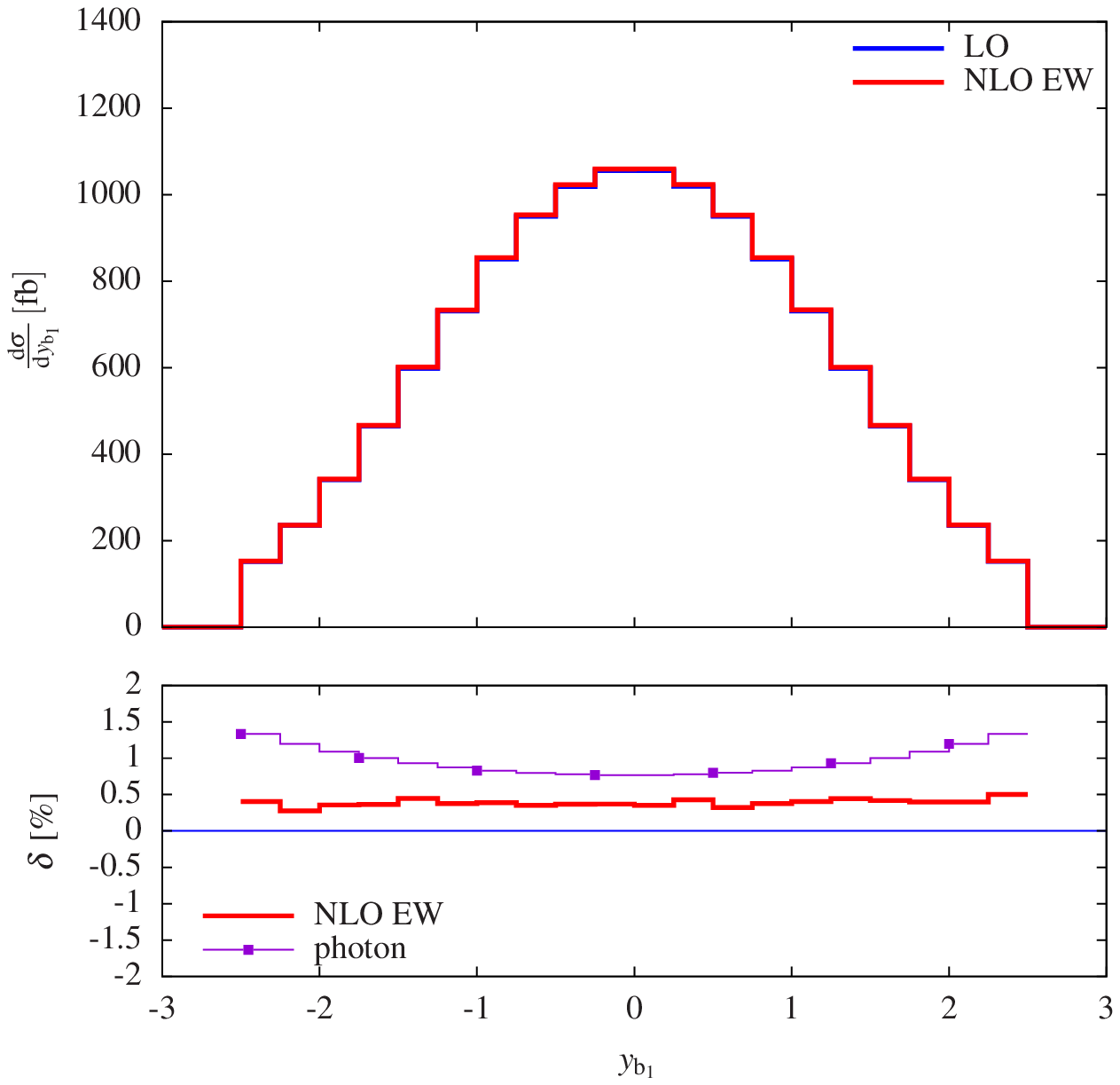}
                \label{plot:rapidity_b1}
        \end{subfigure}
        \hfill
        \begin{subfigure}{0.49\textwidth}
                \subcaption{}
                \includegraphics[width=\textwidth]{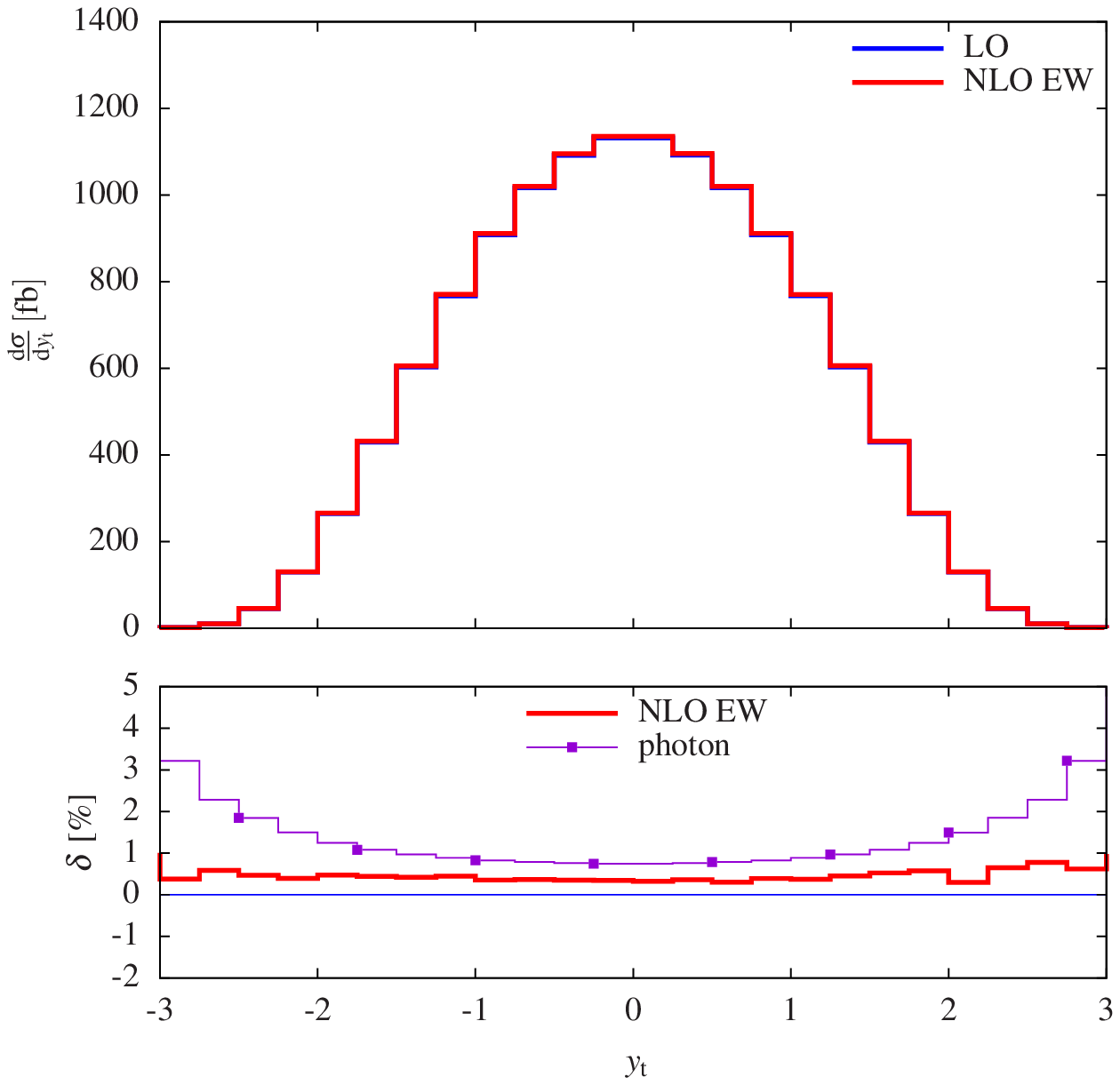}
                \label{plot:rapidity_truth_top}
        \end{subfigure}
        
        \begin{subfigure}{0.49\textwidth}
                \subcaption{}
                \includegraphics[width=\textwidth]{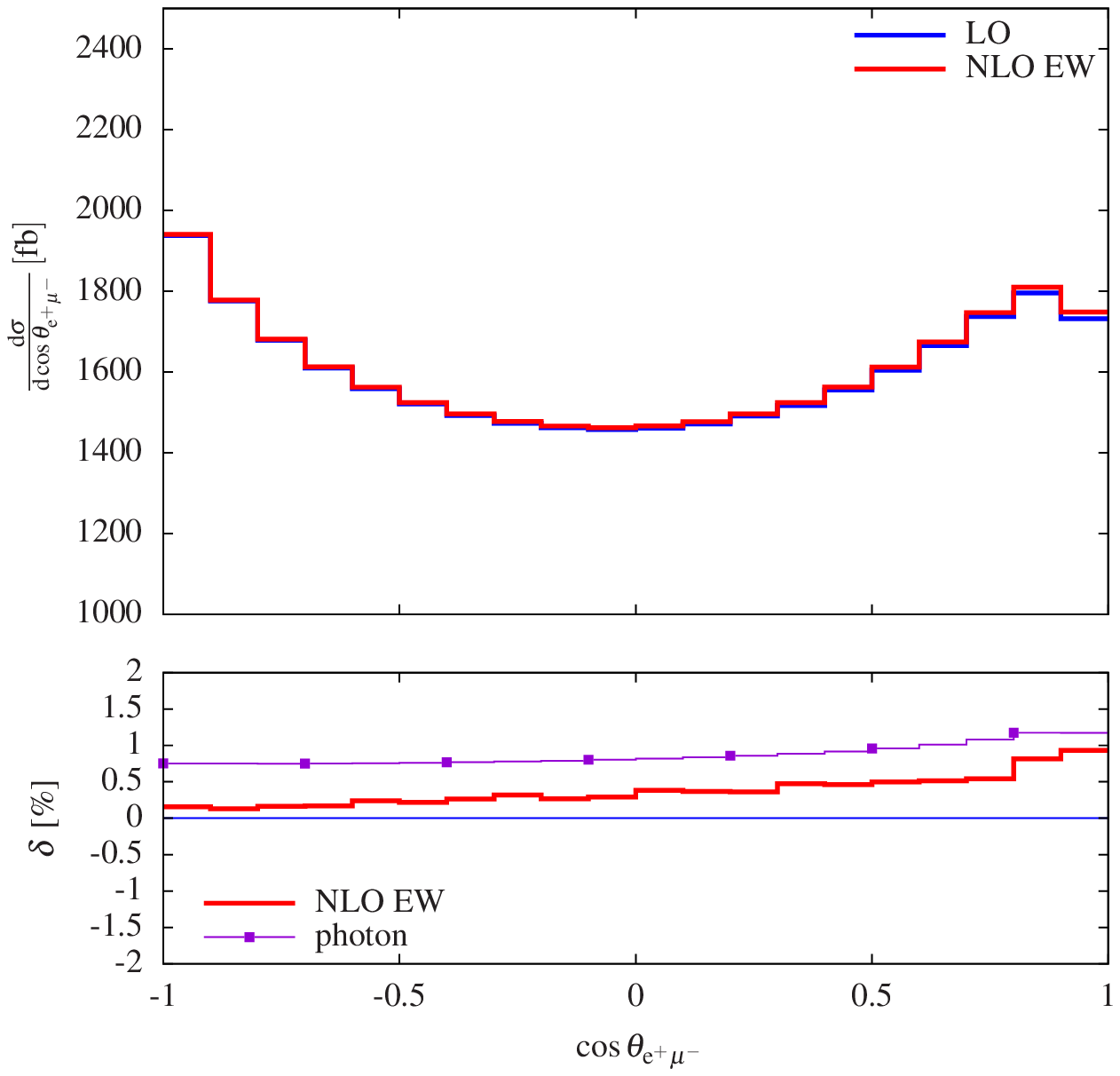}
                \label{plot:cosine_angle_separation_epmu}
        \end{subfigure}
        \hfill
        \begin{subfigure}{0.49\textwidth}
                \subcaption{}
                \includegraphics[width=\textwidth]{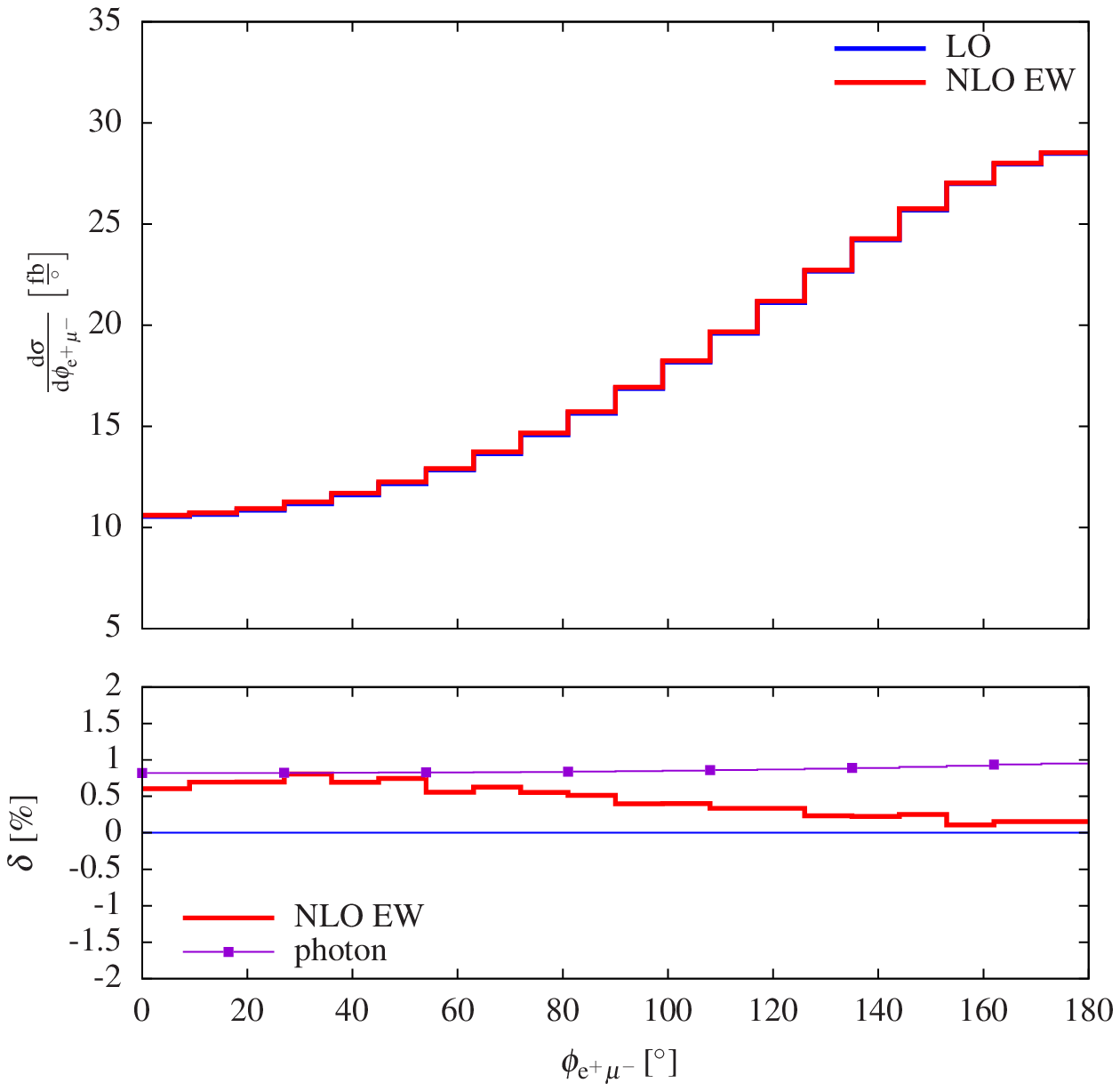}
                \label{plot:azimuthal_angle_separation_epmu}
        \end{subfigure}
        \vspace*{-3ex}
        \caption{\label{fig:further_differential_distributions}%
                Differential distributions at a centre-of-mass energy $\sqrt{s}=13\TeV$ at the LHC: 
                the rapidity of the harder bottom quark \subref{plot:rapidity_b1}~(upper left),
                the rapidity of the reconstructed top quark \subref{plot:rapidity_truth_top}~(upper right), %
                \subref{plot:cosine_angle_separation_epmu} the cosine of the angle between the positron and the muon~(lower left), and %
                \subref{plot:azimuthal_angle_separation_epmu} the azimuthal angle between the positron and the muon in the transverse plane~(lower right).
                The lower panel shows the relative NLO EW correction
                $\delta = \sigma_{\text{NLO EW}} / \sigma_\text{LO} -
                1$ and the relative photon-induced
                contributions $\delta = \sigma_{ \gamma g } /
                \sigma_\text{LO} $ in per cent.  }
\end{figure}%
The rapidity distributions of the harder bottom quark and the
reconstructed top quark are shown in \reffis{plot:rapidity_b1} and
\ref{plot:rapidity_truth_top}, respectively.  The rapidity
distributions of the other final states exhibit flat EW corrections
similar to the ones displayed in \reffi{plot:rapidity_b1}.  Over the
whole rapidity range, the EW corrections are small and do not show any
special features, while the photon-induced contributions are somewhat
more important at high rapidities.  This is particularly true for the
rapidity distribution of the reconstructed top quark.  There, the
photon-induced contribution accounts for up to $3 \%$ for large
rapidities, \ie for top quarks that have been produced close to the
beam, while the EW corrections do not vary over the rapidity range
considered here.  The corrections for the distribution in the cosine
of the angle between the two charged leptons
(\reffi{plot:cosine_angle_separation_epmu}) and the distribution in
the azimuthal angle in the transverse plane between them
(\reffi{plot:azimuthal_angle_separation_epmu}) do not show particular
features and are below $1\%$.

For the observables involving the reconstructed top quarks,
we have found qualitative agreement with the results presented in
\citere{Pagani:2016caq}. Since the calculation of the complete
corrections requires appropriate selection cuts to avoid IR
singularities, no quantitative comparison of distributions is
possible with existing calculations for on-shell top quarks.

\subsection{Comparison to the double-pole approximation}
\label{sec:ComparisonDPA}

We have studied two different DPAs for the off-shell production of
top-quark pairs.  The first one requires two resonant top quarks while
the second one two resonant W bosons.  In this section, we investigate
the quality of these approximations by comparing them with the full
calculation at the cross-section level as well as the
differential-distribution level.

\subsubsection*{Integrated cross section}

We first investigate the DPAs at LO and show results for the total LO
cross section for both channels in
\refta{table:LO_DPA_results_summary}.  While the WW DPA is in
agreement with the full LO result within one per cent, the tt DPA only
agrees within $3 \%$.  This is the order of magnitude $\Gamma/M$
expected for a DPA. The better quality of the WW DPA results from the
fact, that most diagrams for the full process and, in particular,
those with two top resonances contain already two intermediate
W~bosons. On the other hand, there are much more diagrams involving
only one or no resonant top quark.

\begin{table}
\begin{center} 
\begin{tabular}{ c  c  c  c  c}
 Ch. & $\sigma^{\rm WW \; DPA}_{\rm LO}$ [fb] & $\delta^{\rm WW \; DPA}_{\rm LO}$ [$ \% $] & $\sigma^{\rm tt \; DPA}_{\rm LO}$ [fb]  & $\delta^{\rm tt \; DPA}_{\rm LO}$ [$ \% $]\\
  \hline\hline
$\Pg \Pg$ &  $ 2808.4(6)$ & $- 0.56 $ & $ 2738.8(2)  $ & $ -3.0 $  \\
${ q \bar{q}}$      &  $ 372.90(1) $ & $ -0.64 $ & $ 368.82(1) $ & $ -2.2 $ \\ 
  \hline
$\Pp\Pp$         &  $ 3181.3(5)$ & $ -0.57 $ & $3107.6(2)  $ & $ -2.9 $ \\
  \hline
\end{tabular}
\end{center}
        \caption[Comparison of the integrated cross section DPA]{\label{table:LO_DPA_results_summary}
                Integrated LO cross sections for the two DPAs.
                The relative difference is defined as $\delta^{\rm DPA}_{\rm LO} = \sigma^{\rm DPA}_{\rm LO} / \sigma^{\rm Full}_{\rm LO} - 1 $ in per cent.}
\end{table}

At NLO, only the two channels that have been computed in the DPAs are
shown in \refta{table:DPA_results_summary}.  Both approximations
reproduce the total cross section within a per mille.  We recall that
the Born and real matrix elements have been computed with the full
off-shell kinematics.  This is also the case for the contributions
involving the convolution operator ($P$ and $K$ operator in
\citere{Catani:1996vz}), while the one arising from the $I$ operator
has been evaluated with on-shell kinematics applied to the matrix
element featuring two resonant propagators.  As explained before the
factorisable and non-factorisable virtual corrections have been
computed within the DPA.

\begin{table}
\begin{center} 
\begin{tabular}{ c  c  c  c  c}
 Ch. & $\sigma^{\rm WW \; DPA}_{\rm NLO \; EW}$ [fb] & $\delta^{\rm WW\; DPA}_{\rm NLO \; EW}$ [$ \% $] & $\sigma^{\rm tt \; DPA}_{\rm NLO \; EW}$ [fb]  & $\delta^{\rm tt \; DPA}_{\rm NLO \; EW}$ [$ \% $]\\
  \hline\hline
$\Pg \Pg$ &  $ 2832.9(2) $ & $  -0.046  $ & $2836.5(2) $ & $ + 0.082 $  \\
${ q \bar{q}}$      &  $ 377.36(8) $ & $ 0.047 $ & $377.23(5)$ & $ + 0.013 $ \\ 
  \hline
$\Pp\Pp$         &  $3210.5(2) $ & $ -0.037  $ & $ 3214.0(2) $ & $ + 0.072 $ \\
  \hline
\end{tabular}
\end{center}
        \caption[Comparison of the integrated cross section DPA]{\label{table:DPA_results_summary}
                Integrated NLO cross section for the two DPAs.
                Only the channels where the DPAs are applied are shown.
                The relative difference is defined as $\delta^{\rm DPA}_{\rm NLO} = \sigma^{\rm DPA}_{\rm NLO} / \sigma^{\rm Full}_{\rm NLO} - 1 $ in per cent.}
\end{table}

\subsubsection*{Differential distributions}

\begin{figure}
        \setlength{\parskip}{-10pt}
        \begin{subfigure}{0.49\textwidth}
                \subcaption{}
                \includegraphics[width=\textwidth]{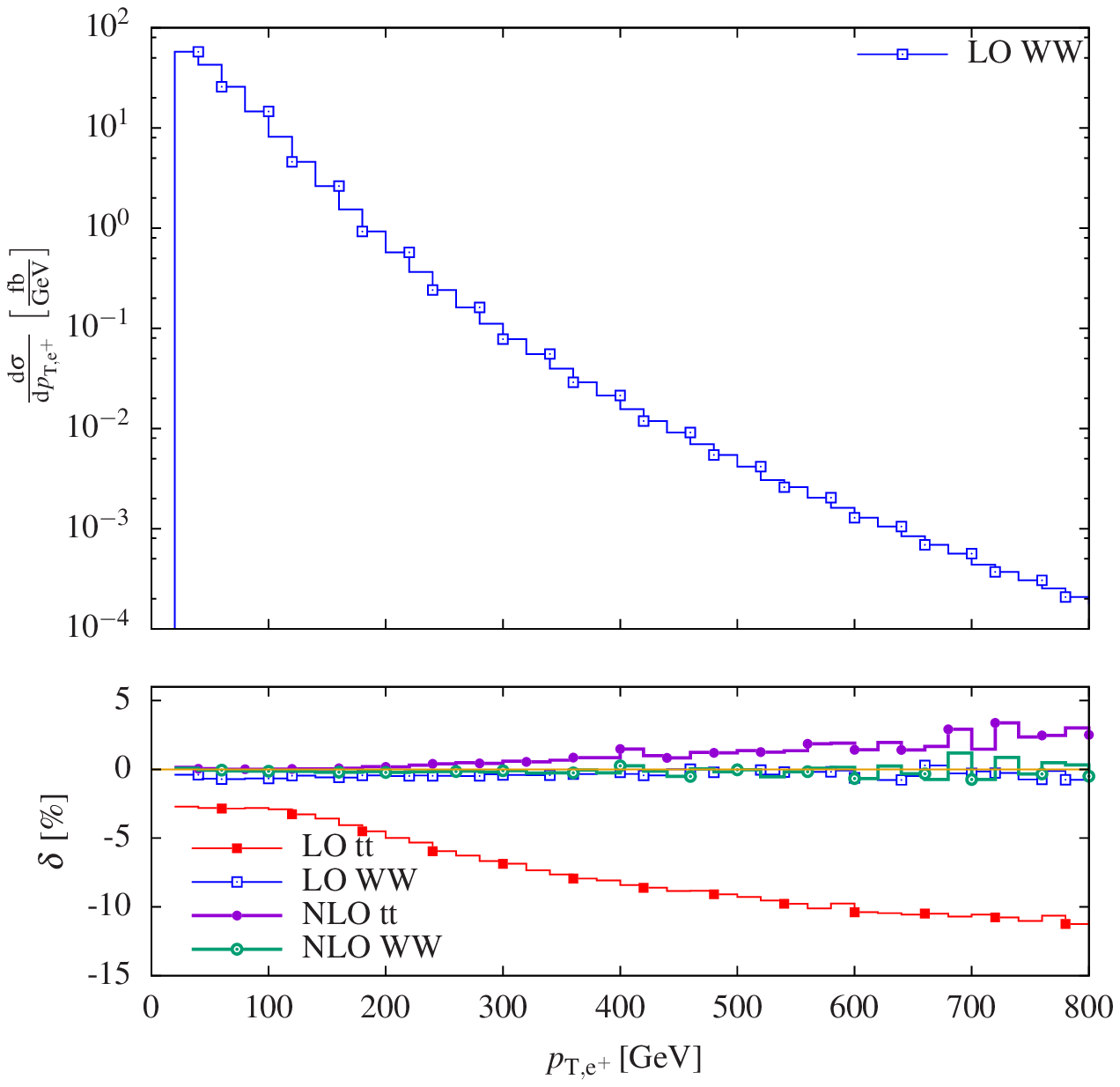}
                \label{plot:transverse_momentum_positron_DPA}
        \end{subfigure}
        \hfill
        \begin{subfigure}{0.49\textwidth}
                \subcaption{}
                \includegraphics[width=\textwidth]{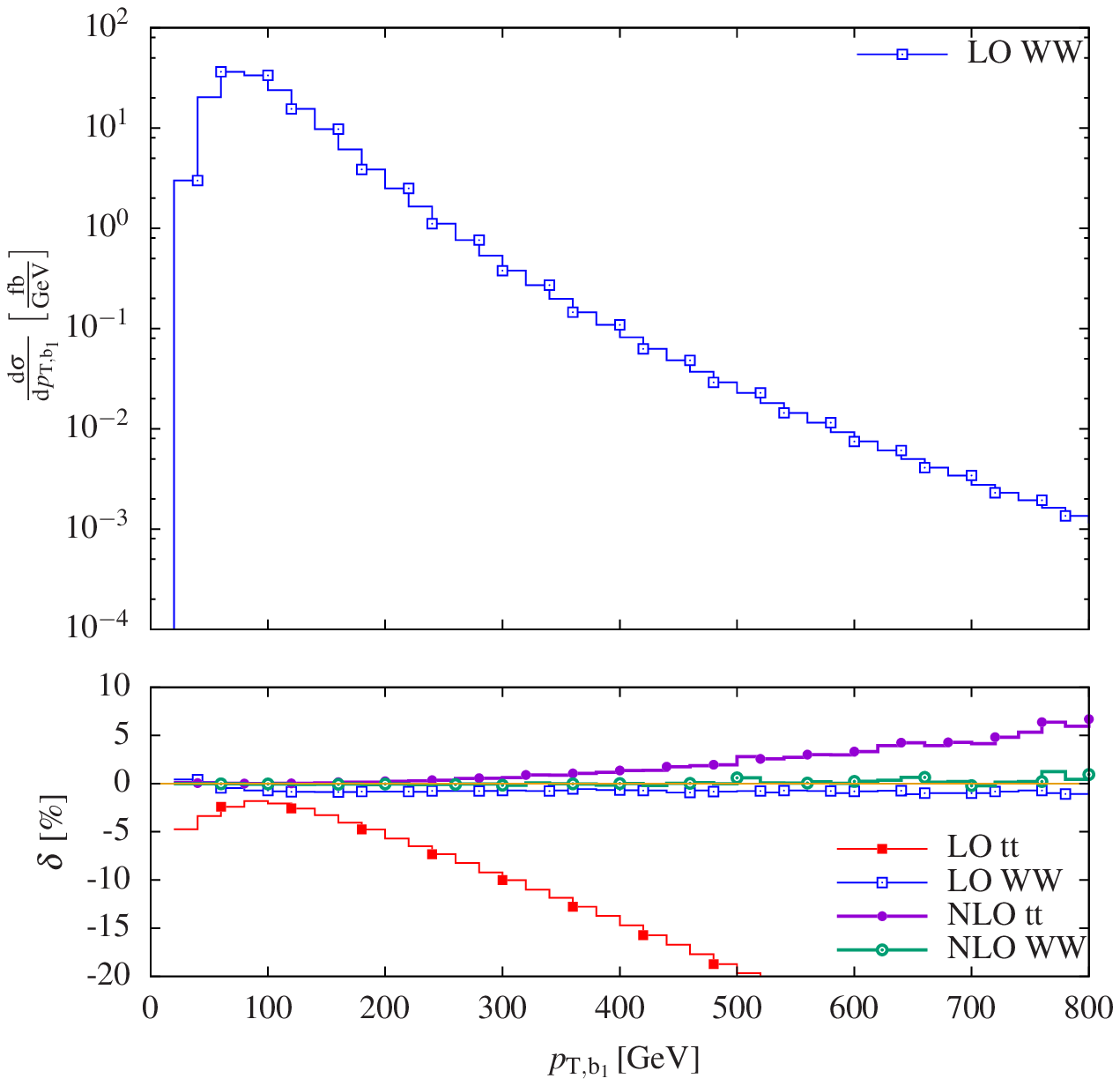}
                \label{plot:transverse_momentum_b1_DPA}
        \end{subfigure}

        \begin{subfigure}{0.49\textwidth}
                \subcaption{}
                \includegraphics[width=\textwidth]{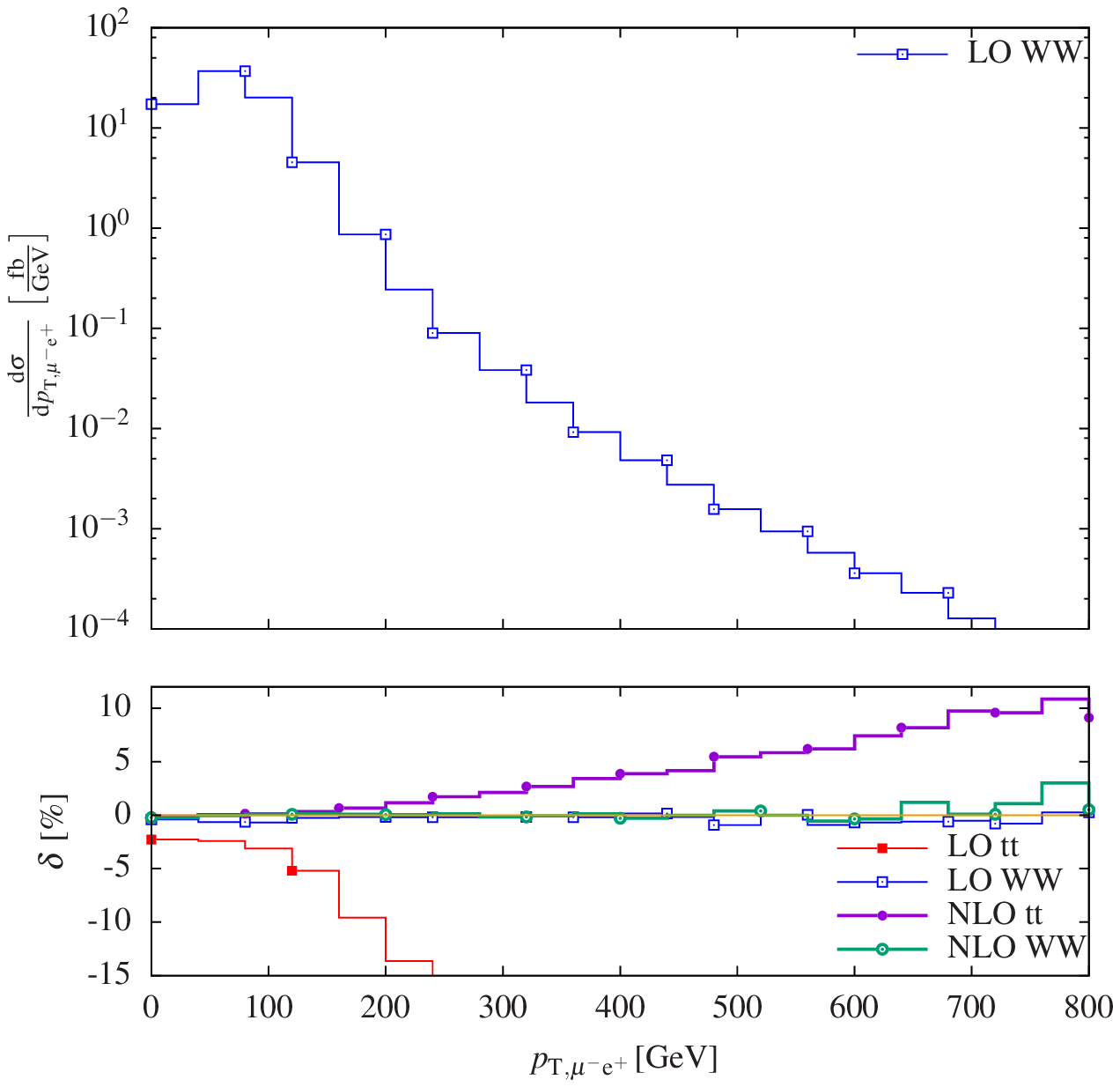}
                \label{plot:transverse_momentum_mupo_DPA}
        \end{subfigure}
        \hfill
        \begin{subfigure}{0.49\textwidth}
                \subcaption{}
                \includegraphics[width=\textwidth]{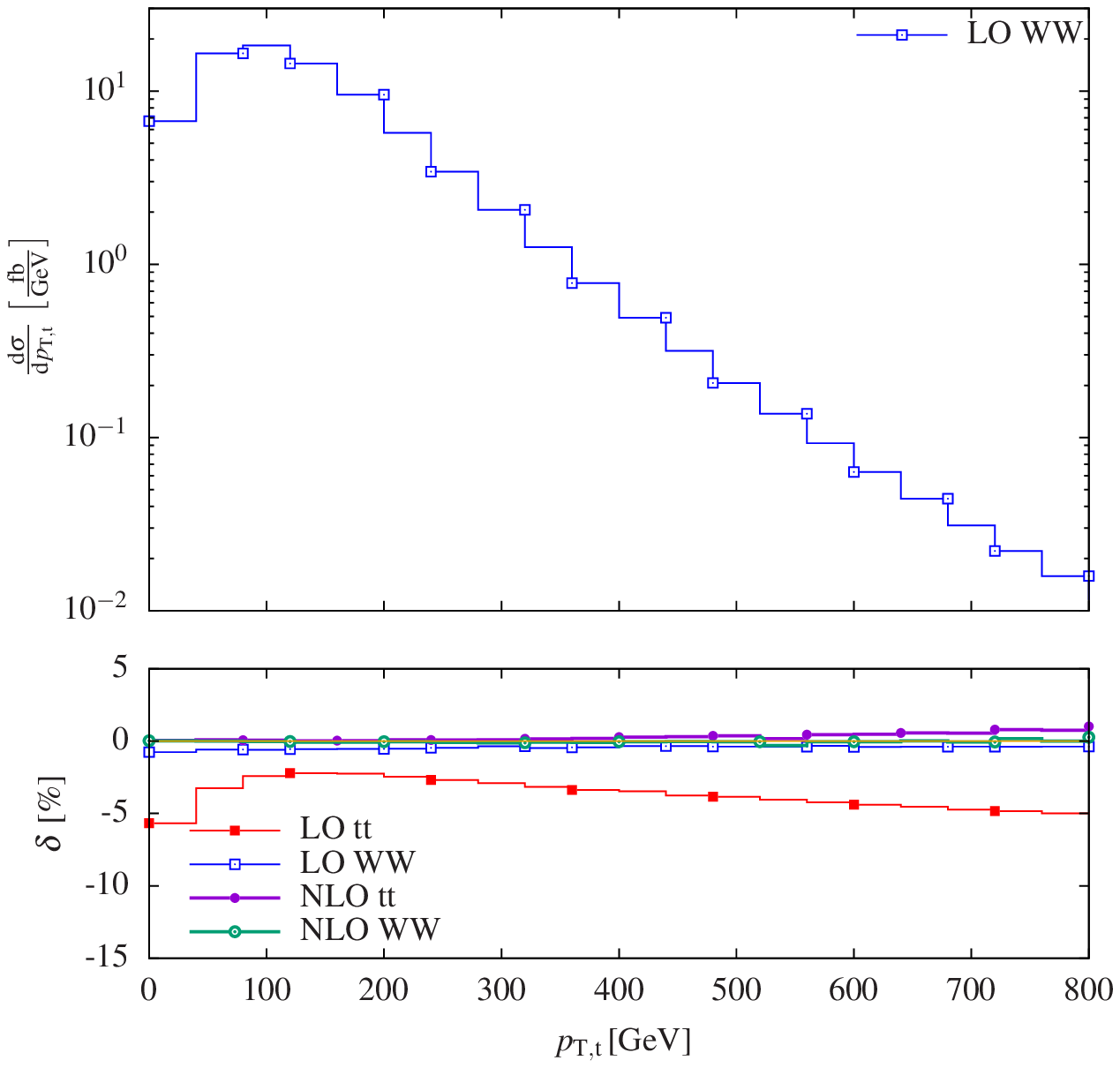}
                \label{plot:transverse_momentum_top_DPA} 
        \end{subfigure}

        \begin{subfigure}{0.49\textwidth}
                \subcaption{}
                \includegraphics[width=\textwidth]{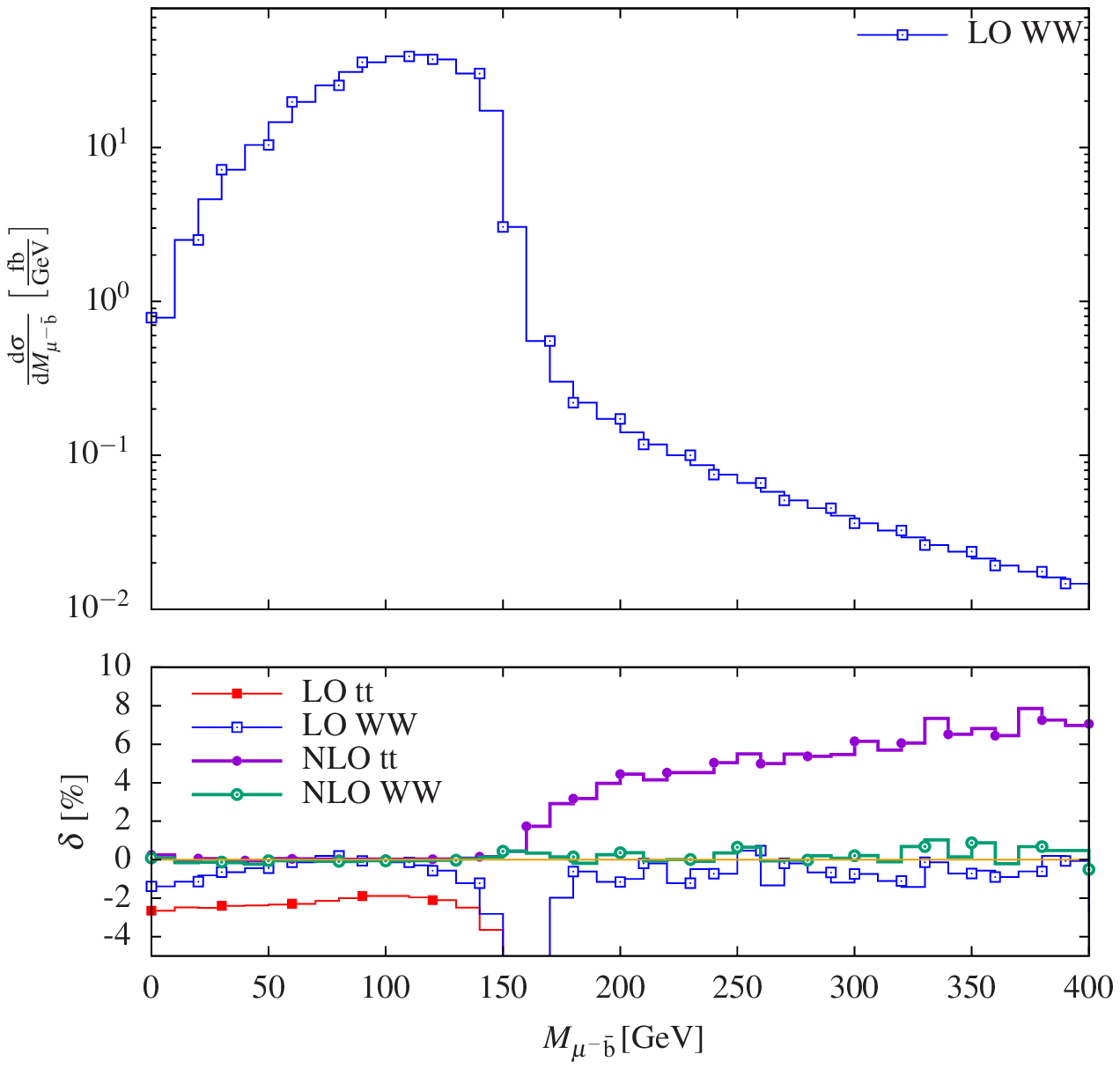}
                \label{plot:invariant_mass_truth_mubx_DPA}
        \end{subfigure}
        \hfill
        \begin{subfigure}{0.49\textwidth}
                \subcaption{}
                \includegraphics[width=\textwidth]{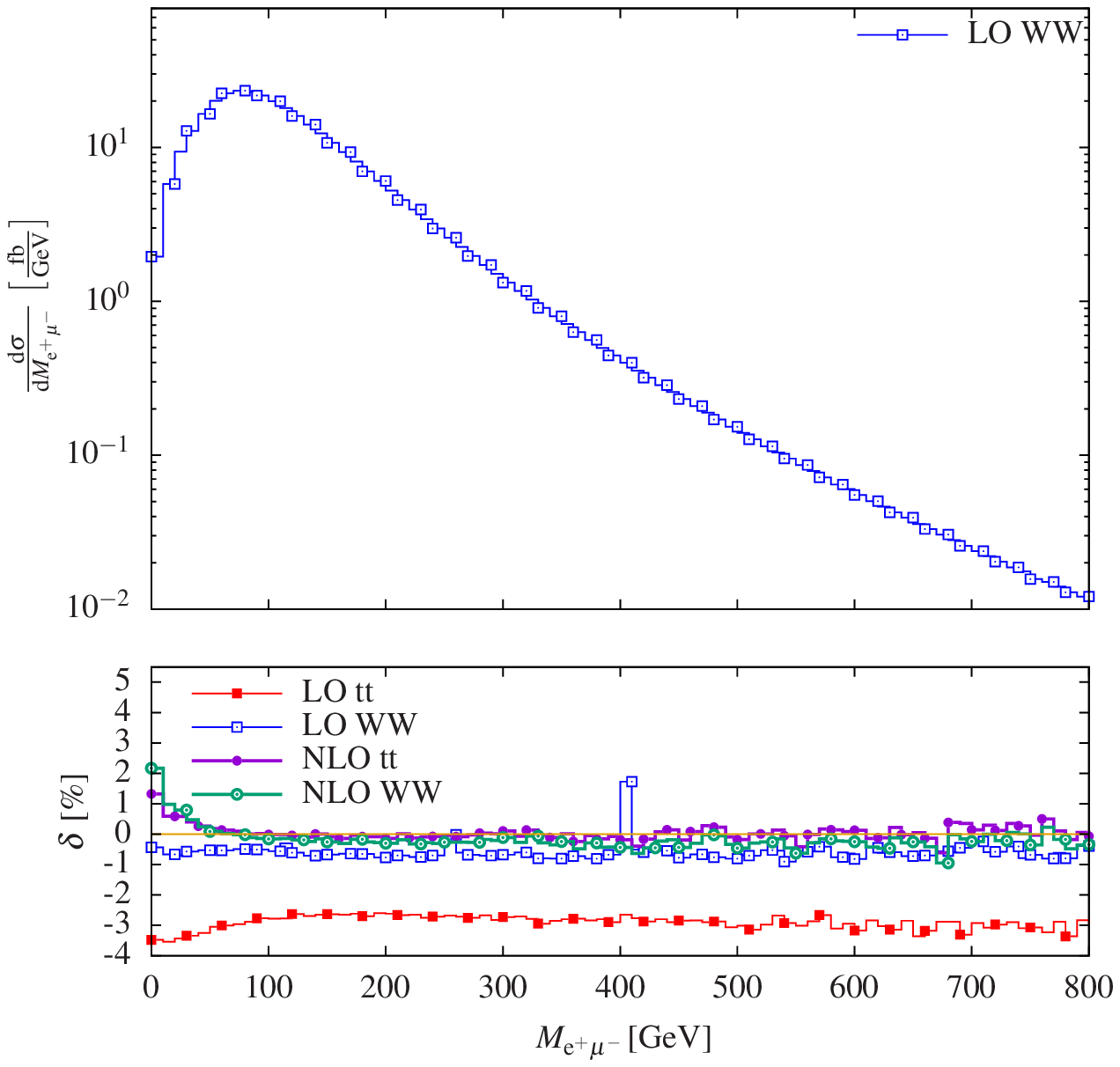}
                \label{plot:invariant_mass_truth_epmu_DPA}
        \end{subfigure}

        \vspace*{-3ex}
        \caption{\label{fig:transverse_momentum_distributions_DPA}%
          Comparison of full calculation and DPAs for various
          distributions at a centre-of-mass energy $\sqrt{s}=13\TeV$
          at the LHC: \subref{plot:transverse_momentum_positron_DPA}
                transverse momentum for the positron~(upper left), %
                \subref{plot:transverse_momentum_b1_DPA} transverse momentum for the harder \Pb~jet~(upper right), %
                \subref{plot:transverse_momentum_mupo_DPA} transverse momentum for the $\mu^-\Pe^+$ system~(middle left) %
                \subref{plot:transverse_momentum_top_DPA} transverse
                momentum  for reconstructed top quark~(middle right), %
                \subref{plot:invariant_mass_truth_mubx_DPA}
                invariant mass for the $\bar{\Pb} \mu^-$
                system~(lower left), and
                \subref{plot:invariant_mass_truth_epmu_DPA}~invariant mass
                for the $\Pe^+ \mu^-$ system~(lower
                right).  In the upper panel the LO distributions for
                the WW DPA are shown.  The lower panel displays the
                relative deviation of the different DPAs from the full
                calculation, $\delta = \sigma_{\text{DPA} } /
                \sigma_\text{Full} - 1$, in per cent.}
\end{figure}%
A comparison of the full calculation with the two DPAs at the
distribution level is presented in
\reffi{fig:transverse_momentum_distributions_DPA}.  The upper panel
contains only one curve (as on the logarithmic scale the three other
curves are indistinguishable) which represents the WW DPA at LO.  In
the NLO computations, the DPA is not applied to the LO contributions,
the real corrections, and to the $P$- and $K$-operator terms.  In the
lower panel, the differences between the approximations and the full
calculation are displayed both at LO and NLO.  The deviation with
respect to the full calculation is defined as $\delta =
\sigma_{\text{DPA} } / \sigma_\text{Full} - 1$ and expressed in per
cent.

The transverse momentum distributions of the electron
(\reffi{plot:transverse_momentum_positron_DPA}), of the harder bottom
jet (\reffi{plot:transverse_momentum_b1_DPA}), and of the $\Pe^+
\mu^-$ system (\reffi{plot:transverse_momentum_mupo_DPA}) display
similar features at LO and NLO for both approximations.  The WW DPA
constitutes a better approximation than the tt one both at LO and NLO
and agrees within $1 \%$ for the observables studied in the considered
phase space.  The tt DPA, on the other hand, deviates by more than $30
\%$ and $11 \%$ at 800\GeV at LO and NLO, respectively.

In the transverse-momentum distributions of the positron and the
harder bottom quark shown in
\reffis{plot:transverse_momentum_positron_DPA} and
\ref{plot:transverse_momentum_b1_DPA} the LO tt DPA deviates from the
full leading order by more than $10\%$ and $20\%$, respectively, for
transverse momenta above 500\GeV. This is due to the fact that it is
easier to produce a particle with large transverse momentum directly
than through an intermediate massive top quark. The effect is smaller
for $p_{\mathrm{T},\Pe^+}$ since there are only very few background
diagrams where the positron does not result from the decay of a
W~boson. This effect is suppressed for the tt DPA at NLO, where the LO
is treated exactly, but still leads to a disagreement of $3\%$ and $6
\%$ for $p_{\mathrm{T},\Pe^+}=800\GeV$ and
$p_{\mathrm{T},\Pb_1}=800\GeV$, respectively.  On the other hand, the
WW DPA approximation describes the full calculation within $1 \%$ over
the full kinematic range displayed.
  
The effects are even more dramatic for the distribution in the
transverse momentum of the muon--positron system shown in
\reffi{plot:transverse_momentum_mupo_DPA}. The cross section is
dominated by events where a pair of top quarks is produced with a
back-to-back kinematics.  For such events, the transverse momentum of
a pair of decay products from different top quarks (for example the
$\mu^-{\rm e}^+$ or the ${\rm b} \overline{{\rm b}}$ pair) tends to be
small, and the high transverse-momentum region in these distributions
receives sizeable corrections from contributions that do not result
from the production of an on-shell top-quark pair.  This explains the
large discrepancy between the tt DPA and the full calculation that
amounts to $11\%$ at NLO and more than $35\%$ at LO for
$p_{\mathrm{T},\mu^-\Pe^+}=800\GeV$. The WW DPA, on the other hand,
allows also contributions with only one or no resonant top quark and
provides a good approximation also for this distribution.

We display in \reffi{plot:transverse_momentum_top_DPA}, the
distribution in the transverse momentum of the reconstructed top
quark.  There the two DPAs agree within $1 \%$ with respect to the
full calculation at NLO.  At LO, the WW DPA works within $1 \%$, while
the tt DPA deviates by up to $5 \%$, which is more or less within the
expected accuracy of a pole approximation.

The invariant-mass distribution of the $\mu^- \bar{\Pb}$ system in
\reffi{plot:invariant_mass_truth_mubx_DPA} displays interesting
features.  Above the threshold at $M^2_{\rm t} - M^2_{\rm W} \simeq
(154\GeV)^2$ the tt DPA is completely off at LO and only agrees within
$10\%$ at NLO. This is due to the fact that this kinematical region is
forbidden for on-shell top quarks and W bosons.  Demanding only
on-shell top quarks, the situation is quite similar as most off-shell
W bosons are close to their mass shell.  Requiring only on-shell
W~bosons, the top-quark invariant mass can become large and allows for
a tail similar as for off-shell W~bosons.  This explains why almost no
deviation from the full calculation is observed above the $M^2_{\rm t}
- M^2_{\rm W}$ threshold for the WW DPA. The large differences of the
WW DPA just above the threshold results from the fact that the
approximation decreases faster than the full cross section owing to
the broadening due to the W-boson width.

For the distribution in the invariant mass of the ${\rm \mu^- e^+}$
system, both approximations reproduce the full calculation at LO and
NLO in shape well.  The difference in the normalisation is as for the
total cross-section (see \refta{table:LO_DPA_results_summary}).

Similarly, rapidity distributions do not show any shape deviation
between neither of the two DPAs and the full calculation.  
The deviation in shape stays below one per cent 
for the distributions in the azimuthal-angle separation and
the cosine of the angle between the two leptons.

To conclude, depending on the considered distribution the tt DPA does
not always describe the full calculation properly.  In some parts of
phase space (especially in the high-energy limit) and for various
distributions the disagreement can reach $10 \%$.  On the other hand,
for all distributions that we have studied the WW DPA describes the
full calculation within a per cent over the considered phase-space
range.  Note that we have specifically checked the transverse-momentum
distribution of the $\Pe^+ \mu^-$ system (which is expected to be most
sensitive to discrepancies between the WW DPA and full calculation)
above 800\GeV and did not find larger deviations of the WW DPA from
the full calculation.  This can be explained by the fact that the WW
DPA features all contributions with single or doubly top resonances
and, thus, the neglected contributions are sub-dominant.

\section{Conclusions}
\label{sec:Conclusions}

For the first time, the production of off-shell top-quark pairs
including their leptonic decays has been computed at the NLO
electroweak level.  In this calculation, all off-shell, non-resonant,
and interference effects have been taken into account.  Moreover, the
photon-induced channels have been evaluated for reference.
The full NLO results have been supplemented by two different
double-pole approximations, one assuming two resonant top quarks and
one requiring two resonant W bosons.

We find electroweak corrections below one per cent for the integrated
cross section, while the contribution from the photon-induced channel
is at the per-cent level.  For differential distributions the
inclusion of electroweak corrections becomes particularly important as
they can account for up to $15\%$ of the leading order.  In this
respect the photon-induced corrections have an effect opposite to the
genuine electroweak corrections.  While the electroweak corrections
are negative in the high-energy limit due to the appearance of Sudakov
logarithms, the photon-induced contributions are positive and increase
with energy.  Nonetheless, in the high-energy region the electroweak
corrections become dominant and account for a significant decrease of
the differential distributions.

We have found that the double-pole approximation requiring two
resonant W~bosons describes the full calculation satisfactorily in the
considered phase-space regions.  On the other hand, we observe
sizeable discrepancies with respect to the full result for the
double-pole approximation requiring two resonant top quarks in several
distributions at both LO and NLO.  This breakdown typically happens in
distributions that involve the decay products of both the top and
antitop quark.  More precisely, differences appear in regions, where
the contributions of two on-shell top quarks are suppressed.  While
such contributions are not taken into account in the top--antitop
double-pole approximation, they are included in the WW one.  We have
found that the WW double-pole approximation constitutes a very good
approximation of the full calculation for all the distributions that
we have investigated.  Nonetheless, it could fail for specific
observables
where off-shell W bosons play an important role.  Thus,
for arbitrary distributions over the whole phase space, one should
only rely on the full calculation.

On the technical side, this calculation demonstrates the ability of
the matrix-element generator \recola and of the integral library
\collier to supply in an efficient and reliable way tree-level and
one-loop amplitudes for complicated processes.

This study provides for the first time the electroweak corrections for
a realistic off-shell production of top quark pairs at the LHC.  It
will help the experimental collaborations to measure the production of
top-quark pairs to even higher precision at the LHC.  Also, the
higher-order corrections described in this article, as electroweak
corrections in general, are relevant for the Standard Model background
of new-physics searches.  Indeed, they grow large exactly in the same
phase-space region where one would expect new-physics contribution to
appear, \ie in the high-energy limit.  Thus, our results will allow to
test the Standard Model with better accuracy and help to discover new
phenomena.

\acknowledgments 
We thank Benedikt Biedermann, Stefan Dittmaier,
Robert Feger, Alexander Huss, Jean-Nicolas Lang, Sandro Uccirati, and
Maria Ubiali for useful discussions.  Specifically, we are grateful to
Robert Feger, Jean-Nicolas Lang, and Sandro Uccirati for providing and
supporting the codes \mocanlo and \recola.  This work was supported by
the Bundesministerium f\"ur Bildung und Forschung (BMBF) under
contract no. 05H15WWCA1.

\appendix

\section{Appendix}
\label{sec:Appendix}

In this appendix we give the explicit expression of the $\Delta$s used
in the computation of the non-factorisable corrections
\eqref{eq:deltanf}--\eqref{eq:delta_i_nf} expressed in
terms of scalar integrals. We simply reproduce the formula of
\citere{Dittmaier:2015bfe} for completeness.  
The functions for the
non-manifestly non-factorisable corrections read:
\begin{align}
        \Delta_{\mathrm{mm'}} \left(i,j\right) &{} \sim
        \begin{aligned}[t]
                 &  - \left( \overline{\overline{s}}_{ij} - M^2_i - M^2_j \right) \bigg\{ C_0 \left( \overline{k}^2_i, \overline{\overline{s}}_{ij}, \overline{k}^2_j, 0, \overline{M}^2_i, \overline{M}^2_j \right) \nonumber
        \end{aligned} \\
        &\quad 
        \begin{aligned}[t]
                &  -  C_0 \left( {M}^2_i, \overline{\overline{s}}_{ij}, {M}^2_j, m_{\gamma}^2, {M}^2_i, {M}^2_j \right) \bigg\} ,
        \end{aligned} \\
        \Delta'_{\mathrm{mm}} \left(i\right) &{} \sim 2 M_i^2 \left\{ \frac{B_0 \left(\overline{k}^2_i, 0, \overline{M}^2_i \right) - B_0 \left(\overline{M}^2_i, m_{\gamma}^2, \overline{M}^2_i \right) }{K_i} - B'_0 \left({M}^2_i, m_{\gamma}^2, {M}^2_i \right) \right\} , \\
        \Delta'_{\mathrm{mf}} \left(i,a\right) &{} \sim
        \begin{aligned}[t]
                 &  - \left( \widetilde{s}_{ia} - M^2_i - m^2_a \right) \bigg\{ C_0 \left( \overline{k}^2_i, {\widetilde{s}}_{ia}, m_a^2, 0, \overline{M}^2_i, m_a^2 \right) \nonumber
        \end{aligned} \\
        &\quad
        \begin{aligned}[t]
                &  - C_0 \left( M^2_i, {\widetilde{s}}_{ia}, m_a^2, m^2_{\gamma}, {M}^2_i, m_a^2 \right) \bigg\} ,
        \end{aligned} \\
        \Delta_{\mathrm{xm}}\left(i;b\right) &{} \sim 
        \begin{aligned}[t]
                 &  - \left( \overline{s}_{ia} - M^2_i - m^2_b \right) \bigg\{ C_0 \left( \overline{k}^2_i, {\overline{s}}_{ib}, m_b^2, 0, \overline{M}^2_i, m_b^2 \right) \nonumber
        \end{aligned} \\
        &\quad 
        \begin{aligned}[t]
                &  - C_0 \left( M^2_i, {\overline{s}}_{ib}, m_b^2, m^2_{\gamma}, {M}^2_i, m_b^2 \right) \bigg\} .
        \end{aligned}
\end{align}
The $\sim$ sign implies that the on-shell limit is taken everywhere
where possible.  This means that all quantities are evaluated
with on-shell kinematics, while only the momenta of the resonant
particles are kept off the mass shell.  Note that each contribution
consists of a scalar integral calculated with complex masses of the
resonances subtracted with the corresponding integral for real masses
but with a photon mass to regularise the IR singularities. While the
IR singularities of the subtracted parts cancel exactly the matching
contributions in the factorisable corrections, those in the original
expressions appear as logarithms of the off-shell propagators and
cancel implicitly upon adding the real corrections.

Finally, the functions for the manifestly non-factorisable virtual
corrections read:
\begin{align}
        \Delta_{\mathrm{ff'}} \left(i, a; j, b\right) & \sim
        \begin{aligned}[t]
                &  - \left( s_{a b} - m^2_a - m^2_b \right) K_i K_j E_0 \left( k_a, \overline{k}_i, -\overline{k}_j, -{k}_b, m_{\gamma}^2, m_a^2, \overline{M}^2_i, \overline{M}^2_j, m_b^2 \right) \nonumber
        \end{aligned} \\
        & \sim
        \begin{aligned}[t]
                &  - \left( s_{a b} - m^2_a - m^2_b \right) K_i K_j \\
                & \times E_0 \left( k^2_a,  \widetilde{s}_{ia},
                  \overline{\overline{s}}_{ij}, \widetilde{s}_{jb},
                  k^2_b, \overline{k}^2_i, \overline{s}_{ja},
                  \overline{s}_{ib}, \overline{k}^2_j, s_{ab},
                  m_{\gamma}^2, m_a^2, \overline{M}^2_i,
                  \overline{M}^2_j, m_b^2 \right) ,
        \end{aligned} \nonumber\\[-1ex]  \\
        \Delta'_{\mathrm{mf'}} \left(i;j,b\right)  & \sim  
        \begin{aligned}[t]
                & - \left( \overline{s}_{i b} - M_i^2 - m^2_b \right) K_j D_0 \left( \overline{k}_i, -\overline{k}_j, -{k}_b, m_{\gamma}^2, \overline{M}^2_i, \overline{M}^2_j, m_b^2 \right) \nonumber
        \end{aligned} \\
        & \sim  
        \begin{aligned}[t]
                & - \left( \overline{s}_{i b} - M_i^2 - m^2_b \right) K_j D_0 \left( \overline{k}^2_i, \overline{\overline{s}}_{ij}, \widetilde{s}_{jb}, k^2_b, \overline{k}^2_j, \overline{s}_{ib}, m_{\gamma}^2, \overline{M}^2_i, \overline{M}^2_j, m_b^2 \right) ,
        \end{aligned} \\
        \Delta_{\mathrm{xf}} \left(i, a; b\right)  & \sim  
        \begin{aligned}[t]
                &  - \left( s_{a b} - m^2_a - m^2_b \right) K_i D_0 \left( {k}_a, \overline{k}_i, -{k}_b, m_{\gamma}^2, m_a^2, \overline{M}^2_i, m_b^2 \right) \nonumber
        \end{aligned} \\
         & \sim  
        \begin{aligned}[t]
                &  - \left( s_{a b} - m^2_a - m^2_b \right) K_i D_0 \left( {k}^2_a, \widetilde{s}_{ia}, \overline{s}_{ib}, k^2_b, \overline{k}_i^2, s_{ab}, m_{\gamma}^2, m_a^2, \overline{M}^2_i, m_b^2 \right),
        \end{aligned}
\end{align}
where the arguments of the scalar integrals have been rewritten in
terms of invariants.  The identification with scalar integrals in terms of momentum arguments reads:
\begin{multline}
D_0 \left( p_1, p_2, p_3, m_{\gamma}^2, m_1^2, m_2^2, m_3^2 \right) \equiv \\
D_0 \left(p_1^2, \left(p_2-p_1\right)^2, \left(p_3-p_2\right)^2, p^2_3, p^2_2, \left(p_3-p_1\right)^2 , m_{\gamma}^2, m_1^2, m_2^2, m_3^2 \right) ,
\end{multline}
and
\begin{multline}
E_0 \left( p_1, p_2, p_3, p_4, m_{\gamma}^2, m_1^2, m_2^2, m_3^2 , m_4^2 \right) \equiv \\
E_0 \big(p_1^2, \left(p_2-p_1\right)^2, \left(p_3-p_2\right)^2, \left(p_4-p_3\right)^2, p^2_4, p^2_2, \\
\left(p_3-p_1\right)^2 , \left(p_4-p_2\right)^2, p^2_3, \left(p_4-p_1\right)^2, m_{\gamma}^2, m_1^2, m_2^2, m_3^2, m_4^2 \big) .
\end{multline}
The scalar integrals used for the numerical evaluation have been obtained from the \collier library \cite{Denner:2014gla,Denner:2016kdg}.

\bibliographystyle{JHEPmod}
\bibliography{ttx_nlo} 

\end{document}

%% file: macros.tex
\newcommand{\Pl}{\ell}

\def\reffi#1{\mbox{Figure~\ref{#1}}}
\def\reffis#1{\mbox{Figures~\ref{#1}}}
\def\refta#1{\mbox{Table~\ref{#1}}}

\def\refse#1{\mbox{Section~\ref{#1}}}
\def\refapp#1{\mbox{App.~\ref{#1}}}
\def\citere#1{\mbox{Ref.~\cite{#1}}}
\def\citeres#1{\mbox{Refs.~\cite{#1}}}

\newcommand{\ri}{\mathrm i}

\newcommand{\ie}{\emph{i.e.}\ }

\def\be{\begin{equation}}
\def\ee{\end{equation}}

\newcommand{\PH}{\ensuremath{\text{H}}\xspace}
\newcommand{\Pj}{\ensuremath{\text{j}}\xspace}
\newcommand{\Pp}{\ensuremath{\text{p}}\xspace}
\newcommand{\Pe}{\ensuremath{\text{e}}\xspace}
\newcommand{\Pb}{\ensuremath{\text{b}}\xspace}

\newcommand{\Pt}{\ensuremath{\text{t}}\xspace}
\newcommand{\Pu}{\ensuremath{\text{u}}\xspace}
\newcommand{\Pd}{\ensuremath{\text{d}}\xspace}
\newcommand{\Ps}{\ensuremath{\text{s}}\xspace}
\newcommand{\Pc}{\ensuremath{\text{c}}\xspace}
\newcommand{\Pg}{\ensuremath{\text{g}}\xspace}

\newcommand{\PW}{\ensuremath{\text{W}}\xspace}
\newcommand{\PZ}{\ensuremath{\text{Z}}\xspace}

\newcommand{\Mt}{\ensuremath{m_\Pt}\xspace}

\newcommand{\MWOS}{\ensuremath{M_\PW^\text{OS}}\xspace}
\newcommand{\MW}{\ensuremath{M_\PW}\xspace}
\newcommand{\MZOS}{\ensuremath{M_\PZ^\text{OS}}\xspace}
\newcommand{\MZ}{\ensuremath{M_\PZ}\xspace}

\newcommand{\Gt}{\ensuremath{\Gamma_\Pt}\xspace}

\newcommand{\GZOS}{\ensuremath{\Gamma_\PZ^\text{OS}}\xspace}

\newcommand{\GWOS}{\ensuremath{\Gamma_\PW^\text{OS}}\xspace}

\newcommand{\GeV}{\ensuremath{\,\text{GeV}}\xspace}
\newcommand{\TeV}{\ensuremath{\,\text{TeV}}\xspace}

\newcommand{\alphas}{\ensuremath{\alpha_\text{s}}\xspace}
\newcommand{\order}[1]{\ensuremath{\mathcal{O}{\left(#1\right)}}\xspace}

\newcommand{\abs}[1]{\left|#1\right|}

\newcommand{\GF}{\ensuremath{G_\mu}}

\newcommand{\pt}{\ensuremath{p_\text{T}}\xspace}
\newcommand{\ptsub}[1]{\ensuremath{p_{\text{T},#1}}\xspace}

\renewcommand{\Re}{\mathop{\mathrm{Re}}\nolimits}

\newcommand{\fullProcess  }{\ensuremath{\Pp\Pp\to\Pe^+\nu_\Pe \mu^-\bar{\nu}_\mu\Pb\bar{\Pb}}\xspace}

\newcommand{\recola}{{\sc Recola}\xspace}
\newcommand{\mocanlo}{{\sc MoCaNLO}\xspace}
\newcommand{\collier}{{\sc Collier}\xspace}
\newcommand{\madgraph}{{\sc\small MadGraph5\_aMC@NLO}\xspace}

\newcolumntype{.}{D{.}{.}{-1}}
\newcolumntype{d}[1]{D{.}{.}{#1}}

\renewcommand{\vec}[1]{\mathbf{#1}}
\colorlet{tableoverheadcolor}{gray!37.5}
\colorlet{tableheadcolor}{gray!25}
\colorlet{tablerowcolor}{gray!12.5}

\newlength{\width}
\newlength{\height}

\def\draftdate{\relax}
\def\mda{\relax}
\def\mua{\relax}
\def\mla{\relax}
\def\draft{
\def\thtystars{******************************}
\def\sixtystars{\thtystars\thtystars}
\typeout{}
\typeout{\sixtystars**}
\typeout{* Draft mode!
         For final version remove \protect\draft\space in source file *}
\typeout{\sixtystars**}
\typeout{}
\def\draftdate{\today}
\def\mua{\marginpar[\boldmath\hfil$\uparrow$]%
                   {\boldmath$\uparrow$\hfil}\color{black}%
                    \typeout{marginpar: $\uparrow$}\ignorespaces}
\def\mda{\color{red}\marginpar[\boldmath\hfil$\downarrow$]%
                   {\boldmath$\downarrow$\hfil}%
                    \typeout{marginpar: $\downarrow$}\ignorespaces}
\def\mla{\marginpar[\boldmath\hfil$\rightarrow$]%
                   {\boldmath$\leftarrow $\hfil}%
                    \typeout{marginpar: $\leftrightarrow$}\ignorespaces}
\def\Mua{\marginpar[\boldmath\hfil$\Uparrow$]%
                   {\boldmath$\Uparrow$\hfil}\color{black}%
                    \typeout{marginpar: $\uparrow$}\ignorespaces}
\def\Mda{\color{red}\marginpar[\boldmath\hfil$\Downarrow$]%
                   {\boldmath$\Downarrow$\hfil}%
                    \typeout{marginpar: $\downarrow$}\ignorespaces}
\def\Mla{\marginpar[\boldmath\hfil\textcolor{red}{$\Rightarrow$}]%
                   {\boldmath\textcolor{red}{$\Leftarrow $}\hfil}%
                    \typeout{marginpar: $\leftrightarrow$}\ignorespaces}
\overfullrule 5pt
\oddsidemargin 15mm
\marginparwidth 29mm
}
